\documentclass[pdflatex,sn-nature,iicol]{sn-jnl}

\usepackage{graphicx}%
\usepackage{multirow}%
\usepackage{amsmath,amssymb,amsfonts}%
\usepackage{xcolor}%
\usepackage{colortbl}%
\usepackage{textcomp}%
\usepackage{booktabs}%
\usepackage{array}%
\usepackage{tikz}%
\usetikzlibrary{arrows.meta,positioning,calc,fit,backgrounds}

\unnumbered

\newif\ifcheckmode \checkmodefalse

\newcommand{\nreal}{20}        
\newcommand{\nseq}{6}

\newcommand{\nstr}{N_{\mathrm{str}}} 
\newcommand{\nseqeval}{8}  

\definecolor{qnlteal}{HTML}{1F9E96}
\definecolor{qnldark}{HTML}{0B5C5A}
\definecolor{fixpurple}{HTML}{7B5AA6}
\definecolor{softgrey}{HTML}{F2F2F2}

\begin{document}

\title[Quantum neuroplastic learning]{Quantum neuroplastic learning for time-series processing}

\author[1]{\fnm{Mostafizur Rahaman} \sur{Laskar} }

\affil[1]{\orgdiv{Division of Information Science and Engineering, School of Electrical Engineering and Computer Science}, \orgname{KTH Royal Institute of Technology}, \orgaddress{\city{Stockholm}, \country{Sweden}}\\ E-mail: morl@kth.se}

\abstract{Quantum machine learning uses quantum states and dynamics to represent and process data. In quantum reservoir computing, the physical system is typically fixed and a classical readout is trained. Inspired by the formation, strengthening and pruning of synapses during biological learning, we propose quantum neuroplastic learning (QNL), which reorganizes the couplings among the qubits of a quantum reservoir, much as experience rewires connections among neurons. Eligibility traces guide local edits to couplings, fields, gate angles and evolution time, and an edit is retained when a statistical bound on its improvement exceeds a complexity cost. We evaluate QNL under restarting and online monitored protocols on short- and long-term memory, nonlinear autoregressive processing and forecasting, and channel equalization tasks. In simulation, QNL enhances memory capacity, develops long-term recall and sustains adaptation across multiple tasks, with the largest benefit for tasks that require extended memory. On the IBM superconducting processor \texttt{ibm\_fez}, with six seven-qubit reservoirs running in parallel on 42 qubits (1.28 million shots in total), configurations learned in simulation and frozen before execution reduced the channel-equalization symbol error rate from 9.7\% to 2.2\% and achieved a lower error rate than a matched-budget tuned baseline.}

\keywords{Neuroplasticity, quantum reservoir computing, temporal learning, superconducting qubits, quantum machine learning}

\maketitle

\begin{figure*}[t]
\centering
\begin{tikzpicture}[x=1cm,y=1cm,
  font=\sffamily\fontsize{6.5}{7.6}\selectfont,
  neu/.style={circle, draw=black!70, fill=white, minimum size=3.6mm, inner sep=0pt},
  qb/.style={circle, draw=qnldark, fill=qnlteal!12, minimum size=4.2mm, inner sep=0pt, font=\sffamily\fontsize{5.5}{6}\selectfont},
  box/.style={draw=black!55, rounded corners=2pt, align=center, inner sep=2.5pt},
  qstep/.style={draw=qnldark, fill=qnlteal!8, rounded corners=2pt, align=left, inner sep=2.5pt, text width=25mm, minimum height=9.5mm},
  arr/.style={-{Stealth[length=1.8mm]}, thick, black!70},
  lab/.style={font=\sffamily\bfseries\fontsize{10}{11}\selectfont},
  ttl/.style={font=\sffamily\bfseries\fontsize{7.5}{8.5}\selectfont, anchor=west}
]
\node[lab, anchor=west] at (0,4.65) {a};
\node[ttl] at (0.32,4.65) {Learning reorganizes neural circuits};
\begin{scope}[shift={(0.35,2.55)}]
  \node[neu] (a1) at (0,1.15) {}; \node[neu] (a2) at (0.9,1.45) {}; \node[neu] (a3) at (1.6,0.7) {};
  \node[neu] (a4) at (0.85,0.0) {}; \node[neu] (a5) at (0.05,0.25) {};
  \draw[black!55, line width=0.6pt] (a1)--(a2) (a2)--(a3) (a3)--(a4) (a4)--(a5) (a5)--(a1) (a1)--(a4);
  \node[anchor=north] at (0.8,-0.2) {before};
\end{scope}
\draw[arr] (2.2,3.3) -- node[above]{experience} (3.1,3.3);
\begin{scope}[shift={(3.45,2.55)}]
  \node[neu] (b1) at (0,1.15) {}; \node[neu] (b2) at (0.9,1.45) {}; \node[neu] (b3) at (1.6,0.7) {};
  \node[neu] (b4) at (0.85,0.0) {}; \node[neu] (b5) at (0.05,0.25) {};
  \draw[black!55, line width=0.6pt] (b1)--(b2) (b3)--(b4) (b4)--(b5) (b5)--(b1);
  \draw[qnlteal, line width=2.2pt] (b2)--(b3);
  \draw[qnlteal, line width=1.0pt] (b2)--(b5);
  \draw[red!65!black, line width=0.8pt, dashed] (b1)--(b4);
  \node[red!65!black, font=\sffamily\bfseries\small] at ($(b1)!0.5!(b4)$) {$\times$};
  \node[anchor=north] at (0.8,-0.2) {after};
\end{scope}
\draw[qnlteal, line width=1.0pt] (0.15,1.8) -- (0.55,1.8); \node[anchor=west] at (0.58,1.8) {formation};
\draw[qnlteal, line width=2.2pt] (2.05,1.8) -- (2.45,1.8); \node[anchor=west] at (2.48,1.8) {potentiation};
\draw[red!65!black, line width=0.8pt, dashed] (4.15,1.8) -- (4.55,1.8); \node[anchor=west] at (4.58,1.8) {elimination};
\node[box, fill=softgrey, text width=56mm, anchor=north west, align=left] at (0.1,1.5)
 {\textbf{Design principle.} Synaptic tags and eligibility traces mark candidate changes; a later signal decides which are kept.};

\node[lab, anchor=west] at (6.35,4.65) {b};
\node[ttl] at (6.67,4.65) {Quantum neuroplastic learning (QNL)};
\node[qstep, anchor=north west] (s1) at (6.45,4.35) {\textbf{1 Nominate} $K$ edits $\boldsymbol\theta\!\to\!\boldsymbol\theta_c$ by eligibility $e_\kappa$};
\node[qstep, anchor=north west] (s2) at (9.55,4.35) {\textbf{2 Probe} $\boldsymbol\theta$ and every $\boldsymbol\theta_c$ on $\nstr$ paired fresh streams};
\node[qstep, anchor=north west, text width=27mm] (s3) at (12.65,4.35) {\textbf{3 Admit} the best edit if $\mathrm{LCB}_c-\alpha\,[\Delta N_c]_+>0$};
\draw[arr, qnldark] (s1) -- (s2);
\draw[arr, qnldark] (s2) -- (s3);
\node[box] (in) at (6.85,2.1) {input\\ $s_k$};
\begin{scope}[shift={(9.0,2.1)}]
  \node[qb] (q0) at (90:0.58) {0}; \node[qb] (q1) at (18:0.58) {1}; \node[qb] (q2) at (-54:0.58) {2};
  \node[qb] (q3) at (-126:0.58) {3}; \node[qb] (q4) at (162:0.58) {4};
  \draw[qnldark!80, line width=0.6pt] (q0)--(q1) (q1)--(q2) (q2)--(q3) (q3)--(q4) (q4)--(q0) (q0)--(q2) (q1)--(q3);
  \draw[qnlteal, line width=1.8pt] (q0)--(q3);
\end{scope}
\node[draw=qnldark, rounded corners=3pt, minimum width=19mm, minimum height=19mm] (res) at (9.0,2.1) {};
\node[anchor=north, align=center] at (9.0,1.0) {quantum reservoir $\Phi_{\boldsymbol\theta}$: couplers,\\ fields, gate angles, clock};
\node[box, text width=15mm] (mon) at (11.5,2.1) {measure or reset\\ (RSP, OLP,\\ NISQRC)};
\node[box, text width=15mm] (ro) at (13.85,2.1) {linear readout\\[1pt] $\hat y_k=\mathbf{w}^{\!\top}\mathbf{x}_k$};
\draw[arr] (in) -- (res);
\draw[arr] (res) -- (mon);
\draw[arr] (mon) -- node[above]{$\mathbf{x}_k$} (ro);
\draw[arr] (ro) -- (15.3,2.1) node[right]{$\hat y_k$};
\draw[arr, qnldark, dashed] (14.95,2.62) -- node[right, align=left, text=qnldark]{task\\loss $L$} (14.95,3.38);
\draw[arr, qnldark] (13.15,3.38) |- (9.0,3.22) -- (9.0,3.06);
\node[text=qnldark, anchor=north] at (11.1,3.2) {admitted edit of $\boldsymbol\theta$};

\end{tikzpicture}
\caption{\textbf{Quantum neuroplastic learning.} \textbf{a}, Experience forms, strengthens and eliminates connections in neural circuits; synaptic tags and eligibility traces mark candidate changes, and later signals decide which are consolidated. \textbf{b}, QNL applies this principle to the physical configuration $\boldsymbol\theta$ of a quantum reservoir. In each block, $K$ local edits (adding, removing or reweighting a coupler, or changing a local field, gate angle or evolution time) are nominated with probabilities set by eligibility traces, and the current and candidate configurations are evaluated on the same $\nstr$ fresh input streams. The best edit is admitted when a nominal lower confidence bound (LCB) on its paired loss reduction exceeds the penalty $\alpha$ per added coupler (Supplementary Note~1); the thick teal coupler marks an admitted edit. Qubit number, measurement protocol and readout architecture are fixed, and the readout is refitted for each evaluated configuration.}
\label{fig:concept}
\end{figure*}
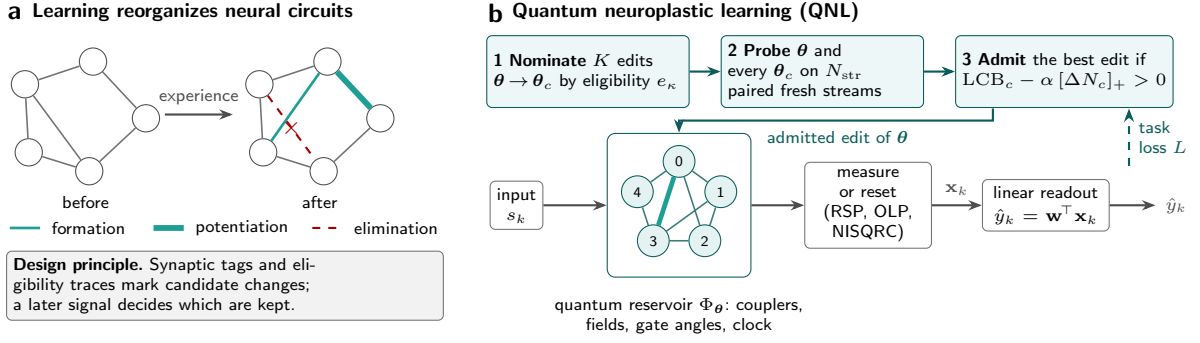

\section*{Introduction}

Quantum machine learning (QML) uses quantum states and dynamics to represent and process data, motivating approaches that can operate within the constraints of noisy quantum hardware\cite{Preskill2018,Biamonte2017,Bharti2022}. Quantum reservoir computing (QRC) provides one such approach for temporal inputs\cite{fujii2017,Nakajima2019,Ghosh2019,Mujal2021}. It inherits the reservoir-computing strategy of driving a fixed dynamical system (called a reservoir) with an input stream and training a linear readout of its high-dimensional state\cite{maass2002,jaeger2004,Lukosevicius2009}. The readout is trained by solving an optimization problem for a fixed internal dynamics that generates high-dimensional features suitable for temporal tasks\cite{Tanaka2019,fujii2017}. The memory and nonlinear processing available to the readout\cite{dambre2012}, however, depend on the reservoir configuration. The exponentially large Hilbert space of a few interacting qubits provides many possible features, but its dimension alone does not determine the learning capacity. In most QRC protocols, the reservoir configuration is drawn once and then left unchanged while the readout is trained. Such protocols perform well on short-term memory (STM) tasks\cite{fujii2017,mujal2023}, relying on the echo-state and fading-memory properties, which make the reservoir state depend on recent inputs and forget its initial condition\cite{jaeger2001echo}, and on the separation of distinct input histories in feature space\cite{maass2002}. Long-term memory and nonlinear time-series prediction have received less attention in the QML literature. 

Neuroplasticity, the capacity of neural circuits to change with experience, underlies learning and memory\cite{Kandel2001,lamprecht2004}. Hebb's postulate that co-activation strengthens connections\cite{Hebb1949} and the discovery of long-term potentiation\cite{Bliss1973} established activity-dependent synaptic change as a substrate of memory. Short-term memory relies on transient, reversible changes in the strength of existing synapses, whereas long-term memory requires changes that are consolidated and stabilized, in part through structural plasticity: the formation, strengthening and elimination of synaptic connections\cite{Kandel2001,lamprecht2004,Holtmaat2009}. Learning a new skill, for example, produces new dendritic spines in the adult cortex, and the small fraction that is stabilized is associated with retention of the skill\cite{Xu2009,Yang2009}. Rewiring sparse connectivity can store more information than changing weights on fixed connections\cite{Chklovskii2004}, and models of synaptic consolidation show that transferring changes from fast to progressively slower variables extends memory lifetimes substantially\cite{fusi2005,benna2016}.
 
Which changes persist is decided by mechanisms that link local activity to later signals. Synaptic tags and eligibility traces mark candidate changes, and delayed neuromodulatory signals determine whether they are consolidated\cite{frey1997,Gerstner2018,Fremaux2016}. These principles have informed artificial learning: networks that grow or rewire their connectivity\cite{Stanley2002,Mocanu2018,Bellec2018}, eligibility-based learning in spiking networks\cite{Bellec2020}, structural learning as stochastic sampling\cite{Kappel2015}, and intrinsic and self-organizing plasticity that improves memory in classical reservoirs\cite{Steil2007,Schrauwen2008,Lazar2009}. Continual learning adds a further requirement, because networks must avoid forgetting earlier tasks\cite{Kirkpatrick2017,Zenke2017} and can lose the ability to learn new ones\cite{Dohare2024}, and transferring biological features to artificial systems requires care in choosing which ones are useful\cite{bikic2025}.
 
Adapting quantum dynamics has been explored in QML. Variational quantum algorithms train circuit parameters, commonly by gradient descent\cite{Cerezo2021}, with limitations from barren plateaus\cite{McClean2018,Larocca2025} and from the measurements needed to estimate gradients. Adaptive ans\"atze and architecture search change circuit structure, usually by offline optimization of a fixed objective\cite{Grimsley2019,Du2022}, and continual learning with unitary variational circuits has addressed catastrophic forgetting\cite{Jiang2022} and loss of plasticity\cite{Chen2026}. In contrast, QRC performs temporal tasks with a fixed reservoir configuration\cite{fujii2017}. This raises the question of whether the physical configuration of a monitored reservoir can be improved during learning by local, gradient-free updates, at fixed qubit number, measurement protocol and readout architecture, and whether such adaptation persists when the task changes. 
 
Here, inspired by structural plasticity, we introduce quantum neuroplastic learning (QNL), a learning principle that reorganizes the physical configuration of a quantum reservoir in response to task performance (Fig.~\ref{fig:concept}). In analogy with fast synaptic changes and slower structural remodelling, QNL combines a readout that is refitted for every evaluated configuration with a reservoir configuration that changes slowly, one admitted edit at a time. QNL proposes local, discrete edits to the active couplers and their strengths, local fields, gate angles or evolution time, which correspond to synaptic formation, elimination and re-weighting. Eligibility traces favour edit types that have recently improved performance, and the best candidate is admitted when a nominal lower confidence bound on its loss reduction, measured on fresh paired input streams, exceeds a penalty for added couplers. Qubit number, measurement protocol and readout architecture are unchanged; the cost is paid in additional evaluations during learning.
 
We evaluate our proposed QNL across reservoir maps, measurement budgets and changing tasks. In monitored erase-and-write reservoirs\cite{mujal2023,MorguiSancho2026}, reorganization enhances STM capacity with exact expectation values and develops long-term recall where fixed configurations retain negligible memory; the gain is smaller under amplitude-damping monitoring and depends on the measurement budget. QNL keeps adapting over sequences of $16$ tasks and improves one-step-ahead forecasting of nonlinear autoregressive series\cite{Atiya2000}. On the NISQRC Hamiltonian ansatz for channel equalization\cite{Hu2024}, QNL reaches error rates comparable to an offline random search with a matched budget. On the IBM processor \texttt{ibm\_fez}, a configuration learned in simulation and frozen before execution reduces the symbol error rate from $9.7\%$ to $2.2\%$ with fewer CZ gates, and in a second, chain-balanced run it has the lowest error rate ($4.2\%$, against $5.8\%$ for a matched-budget tuned baseline and $13.0\%$ for the reference).

\section*{Results}

\subsection*{Quantum neuroplastic learning}

In a quantum reservoir, the configuration, together with the input encoding and the measurement, determines which past inputs remain decodable from the measured features, and a trained readout can use the information that the reservoir retains. Hence, training the readout alone is analogous to plasticity on a fixed connectivity: it changes how the available signals are combined, while the configuration sets which signals persist. QNL introduces a second, slower level of learning that edits the configuration itself, in analogy with the formation, elimination and re-weighting of synapses associated with long-term memory\cite{lamprecht2004,Holtmaat2009} and with the transfer of changes from fast to slow variables in models of memory consolidation\cite{benna2016}.

We consider a quantum reservoir whose physical configuration $\boldsymbol\theta$ collects the active couplers and their strengths, the local fields or single-qubit gate angles, and the evolution time per input. At step $k$, the reservoir updates its state $\rho_k$ and produces the features $\mathbf x_k$ and the output $\hat y_k$ as
\begin{align}
\rho_k&=\Phi_{\boldsymbol\theta}(s_k)\big[\rho_{k-1}\big],\qquad
\mathbf x_k&=\mathcal M(\rho_k),\qquad\nonumber\\
\hat y_k&=\mathbf w^{\!\top}\mathbf x_k,
\label{eq:qrc}
\end{align}
where $s_k$ is the input, $\Phi_{\boldsymbol\theta}(s_k)$ is a completely positive, trace-preserving map that includes any measurement back-action or reset, $\mathcal M$ returns measured expectation values or outcome frequencies, and the readout weights $\mathbf w$ are trained for each task. We denote by $L(\boldsymbol\theta;\xi)$ the task loss on an input stream $\xi$, evaluated on the part of the stream that is not used to train the readout. In conventional QRC, $\boldsymbol\theta$ is fixed and $\mathbf w$ is learned; QNL also updates $\boldsymbol\theta$ in plasticity blocks $b=1,2,\dots$ (Fig.~\ref{fig:concept}b) according to
\begin{align}
\boldsymbol\theta^{(b+1)}&=
\begin{cases}
\boldsymbol\theta_{c^\ast}, & \Gamma\big(\boldsymbol\theta_{c^\ast};\boldsymbol\theta^{(b)}\big)>0,\\
\boldsymbol\theta^{(b)}, & \text{otherwise},
\end{cases}
\qquad\nonumber\\
c^\ast&=\operatorname*{arg\,max}_{1\le c\le K}\,\Gamma\big(\boldsymbol\theta_c;\boldsymbol\theta^{(b)}\big).
\label{eq:qnl}
\end{align}
Here, the $K$ candidates $\boldsymbol\theta_c$ each differ from $\boldsymbol\theta^{(b)}$ by one local edit and are drawn with probabilities biased by eligibility traces over edit types, and $\Gamma$ is an admission margin: a lower confidence bound on the loss reduction of the candidate, measured on fresh input streams shared with the current configuration, minus a cost per added coupler. Thus, an edit is retained when there is statistical evidence that it improves the task loss on unseen data. QNL requires forward evaluations of the task loss, which a reservoir computer measures as part of its normal operation, so it can be applied to physical reservoirs whose dynamics include measurement back-action and sampling noise; once frozen, a QNL configuration is executed with the same protocol as the reservoir from which it was derived. The explicit forms of the nomination rule, of $\Gamma$ and of the eligibility update are given in Supplementary Note~1 (Supplementary Eqs.~(S1)--(S4)), and the settings of each experiment in Supplementary Table~1.

Note that QNL differs from variational quantum algorithms\cite{Cerezo2021}, which optimize the continuous gate parameters of a fixed circuit structure, typically with gradient-based methods. In contrast, QNL reorganizes the reservoir itself through discrete, gradient-free edits of its couplers, fields and evolution time, which are admitted when the paired improvement in the task loss exceeds a statistical lower bound, while the readout remains linear.

\begin{figure*}[t]
\centering
\includegraphics[width=0.77\textwidth]{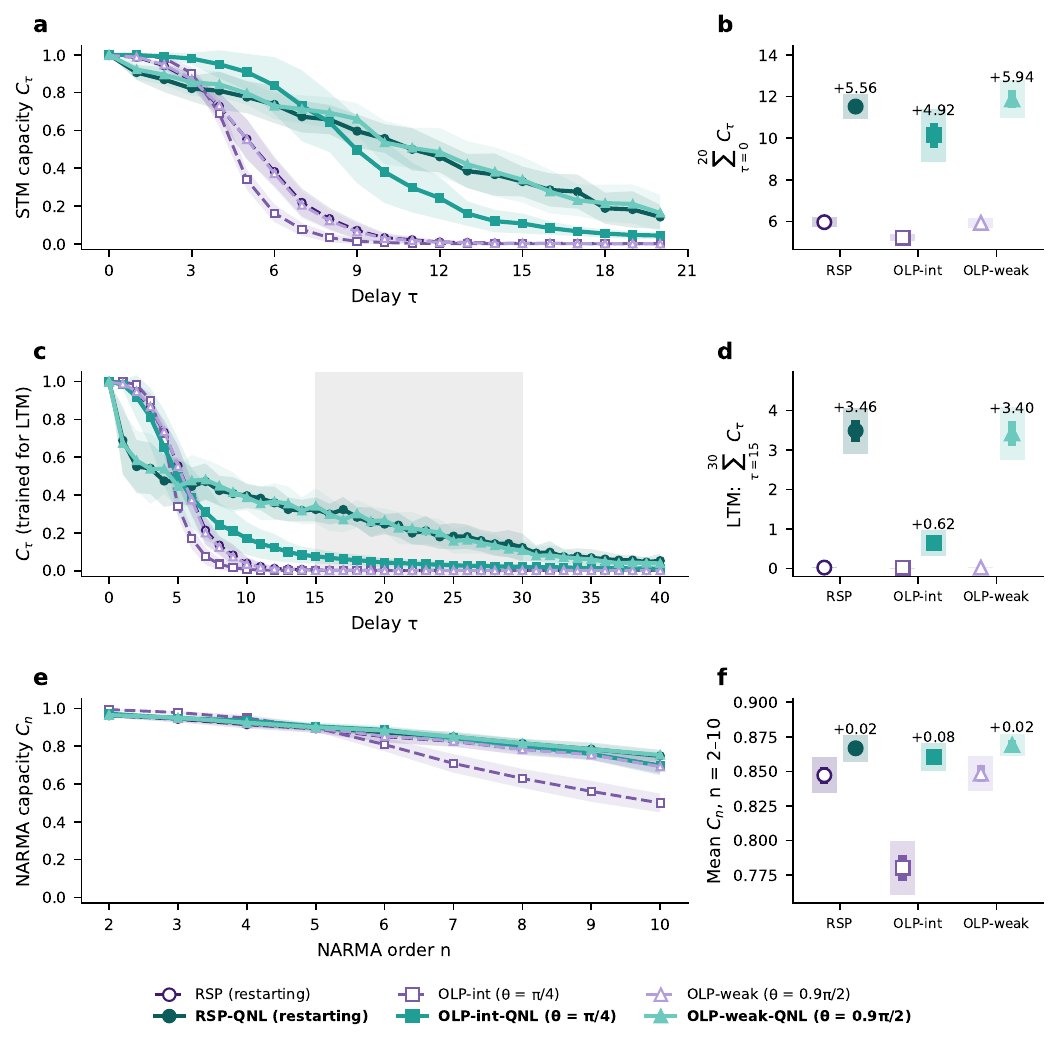}
\caption{\textbf{Plastic reorganization of monitored erase-and-write reservoirs.} Five-qubit reservoirs with the restarting protocol (RSP) and online protocols with intermediate (OLP-int, $\theta=\pi/4$) and weak (OLP-weak, $\theta=0.9\pi/2$) dephasing monitoring, before (purple, dashed, open symbols) and after QNL (teal, solid, filled symbols). Separate learning runs from the same base reservoir optimize short-term memory (STM; $\tau=0$--$20$), long-term memory (LTM; $\tau=15$--$30$) and NARMA processing ($n=2$--$10$). \textbf{a},\textbf{b}, STM capacity $C_\tau$ and summed STM capacity. \textbf{c},\textbf{d}, Capacity profile of the reservoirs trained for LTM (grey band, objective) and LTM capacity $\sum_{\tau=15}^{30}C_\tau$. \textbf{e},\textbf{f}, NARMA capacity $C_n$ and its mean over $n=2$--$10$. Capacities use exact expectation values on held-out streams. Lines and bands, mean $\pm$ s.d. over $n=\nreal$ reservoir realizations; boxes, mean $\pm$ s.d.; thick bars, 95\% Student-$t$ confidence interval; numbers, paired mean gain.}
\label{fig:ew}
\end{figure*}

\subsection*{Plasticity reorganizes memory in monitored reservoirs}

We first apply our proposed QNL to the monitored erase-and-write reservoirs of refs.~\cite{mujal2023,MorguiSancho2026}: five qubits coupled by a fully connected transverse-field Ising Hamiltonian, with the input written into qubit~0 at every step (Methods). Features are read out either with the restarting protocol (RSP), in which the reservoir is re-prepared and the input history replayed for every measurement, or with the online protocol (OLP), in which non-selective dephasing measurements of strength $\theta$ multiply the coherences of each measured qubit by $\sin\theta$. QNL edits the couplers, local fields and clock, while the measurement strength and qubit number remain unchanged.

As shown in Fig.~\ref{fig:ew}a,b, reorganization increases the summed short-term memory capacity by $5.56$ for RSP, $4.92$ for OLP-int (intermediate monitoring strength, $\theta=\pi/4$) and $5.94$ for OLP-weak ($\theta=0.9\pi/2$). In the fixed reservoirs, the capacity falls below $0.1$ within seven to nine steps. After QNL, the RSP and OLP-weak reservoirs still recall the input from $20$ steps earlier with $C_{20}\approx0.15$, at the cost of slightly lower recall at $\tau=1$ ($C_1\approx0.91$ instead of $\approx0.99$). Reservoirs trained for long-term memory (Fig.~\ref{fig:ew}c,d) develop a capacity plateau of $C_\tau\approx0.25$--$0.35$ at $\tau=15$--$20$, where all fixed reservoirs have negligible memory; the LTM capacity increases from near zero to $3.46$ for RSP and $3.40$ for OLP-weak, and to $0.62$ for OLP-int, whose measurements multiply the coherences of every qubit by $\sin(\pi/4)\approx0.71$. Thus, plasticity redistributes memory within the limits set by the measurement protocol. For the NARMA tasks (Fig.~\ref{fig:ew}e,f), the gains in mean capacity are small ($+0.02$ to $+0.08$), because the fixed reservoirs already reach $C_n>0.9$ at low orders, and they are concentrated at high orders, where memory is limiting.

Two further studies test the limits of these gains. Under amplitude-damping monitoring, QNL increases the summed STM capacity by $0.99$ (OLP-int) and $0.43$ (OLP-weak), smaller gains than for the erase-and-write map, which suggests that the monitoring, which QNL does not edit, dominates the memory profile (Supplementary Note~4). When features are estimated from a finite number of measurements, the reservoirs reorganized with exact features exceed the fixed ones from $\sim10^5$ measurements per step with RSP and from $10^8$ with online monitoring, because QNL moves memory to weak feature components that require large ensembles to resolve (Supplementary Note~5). Hence, the admission loss should be estimated at the measurement budget of the intended deployment, as in the NISQRC and hardware experiments below.

\begin{figure*}[t]
\centering
\includegraphics[width=0.92\textwidth]{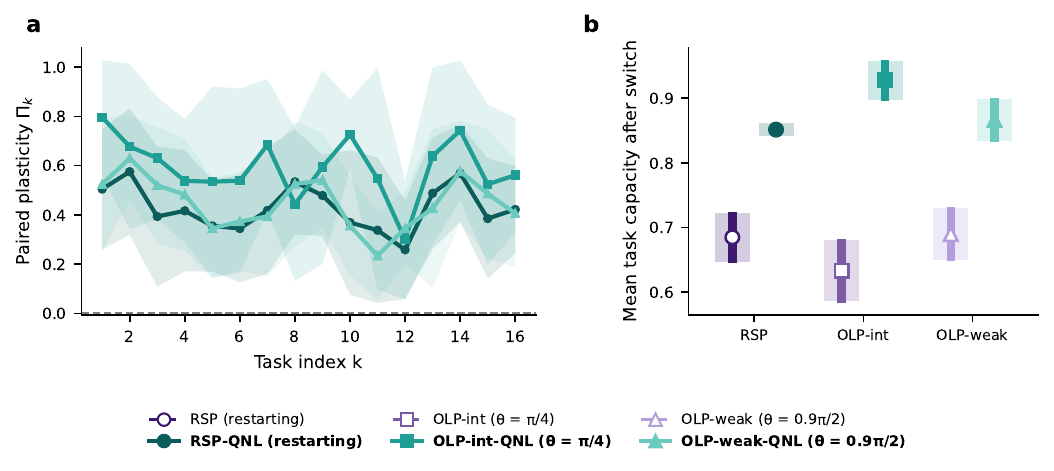}
\caption{\textbf{Adaptation continues across task switches.} Sequences of 16 tasks drawn from four STM windows and three NARMA orders without immediate repetition; after each switch, QNL continues from the configuration reached on the previous task. \textbf{a}, Paired plasticity $\Pi_k$ (Supplementary Eq.~(S5)); the dashed line ($\Pi_k=0$) corresponds to a fixed reservoir. Lines and bands, mean $\pm$ s.d. over $n=\nseq$ sequences. \textbf{b}, Capacity after the switch, $1-L$, averaged over the 16 tasks of each sequence, for fixed reservoirs (purple) and QNL (teal). Boxes, mean $\pm$ s.d.; bars, 95\% Student-$t$ confidence interval.}
\label{fig:plasticity}
\end{figure*}

\subsection*{Adaptation continues across task switches}

A learner that adapts well to one task may lose the ability to adapt to the next\cite{Dohare2024}. To examine this, we run QNL continually through sequences of 16 tasks, switching between memory windows at different delays and NARMA orders (Supplementary Note~6), and quantify each switch by the paired plasticity $\Pi_k$, the relative loss reduction obtained by adapting to task $k$ from the configuration inherited from task $k-1$ (Supplementary Eq.~(S5)); $\Pi_k=0$ for a fixed reservoir. In Fig.~\ref{fig:plasticity}a, the mean $\Pi_k$ remains between $0.2$ and $0.8$ for all 16 tasks and all three protocols, and its fluctuations between tasks are larger than the systematic change over the sequence. Averaged over each sequence, the capacity after the switch is $0.85$, $0.93$ and $0.87$ for RSP-QNL, OLP-int-QNL and OLP-weak-QNL, compared with $0.69$, $0.63$ and $0.69$ for the corresponding fixed reservoirs (Fig.~\ref{fig:plasticity}b). The quantity $\Pi_k$ measures adaptation to the new task; retention of earlier tasks is left for future work.

\begin{figure*}[t]
\centering
\includegraphics[width=0.92\textwidth]{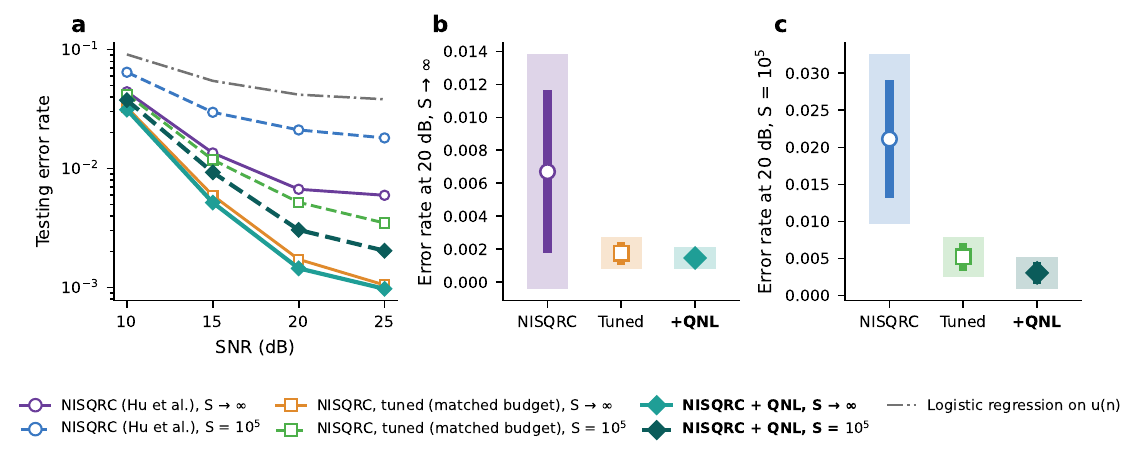}
\caption{\textbf{QNL on top of NISQRC (Hamiltonian ansatz).} Channel equalization with the $(2+4)$-qubit NISQRC of ref.~\cite{Hu2024}: random instances from the authors' parameter distribution (NISQRC; labelled `Hu et al.' in the legend), the same distribution tuned by random search with the same number of candidates (tuned), and NISQRC after QNL (+QNL). Both learning methods use $S=10^5$-shot features at 20\,dB. \textbf{a}, Testing error rate versus signal-to-noise ratio for exact probabilities ($S\to\infty$, solid) and $S=10^5$ (dashed); dash-dotted, logistic regression on the received signal $u(n)$. \textbf{b},\textbf{c}, Error rate at 20\,dB for $S\to\infty$ and $S=10^5$. Boxes, mean $\pm$ s.d. over $n=8$ realizations; bars, approximate 95\% confidence interval.}
\label{fig:nisqrc}
\end{figure*}

\subsection*{Plasticity on top of NISQRC}

Next, we apply QNL to NISQRC\cite{Hu2024}, in which the readout qubits of an $(M+R)$-qubit reservoir are measured and deterministically reset after every input, while the memory qubits carry information forward. We use the $(2+4)$-qubit Hamiltonian ansatz and the channel-equalization task of ref.~\cite{Hu2024}, in which four-level symbols, distorted by a linear channel with an eight-symbol impulse response, a polynomial nonlinearity and Gaussian noise, are to be recovered (Methods). QNL learns at 20\,dB with features sampled at $S=10^5$ shots. As an offline control, a tuned baseline evaluates the same number of candidate configurations (80) on the same streams and shot budget and keeps the best one; QNL additionally re-evaluates its current configuration in every block, which corresponds to 25\% more stream evaluations.

At 20\,dB, QNL reduces the error rate of the random NISQRC instances from $6.7\times10^{-3}$ to $1.5\times10^{-3}$ for exact probabilities and from $2.1\times10^{-2}$ to $3.1\times10^{-3}$ at $S=10^5$ (Fig.~\ref{fig:nisqrc}). The tuned baseline reaches $1.7\times10^{-3}$ and $5.2\times10^{-3}$, close to the values reported in ref.~\cite{Hu2024}, and with $n=8$ realizations its differences from QNL lie within the spread across realizations. From $S\to\infty$ to $S=10^5$, the relative increase in the error rate is smaller for QNL than for the random and tuned reservoirs, consistent with admission based on shot-sampled losses. The two approaches reach comparable error rates; the difference is that QNL improves the configuration in use incrementally.

\begin{figure*}[t]
\centering
\includegraphics[width=0.92\textwidth]{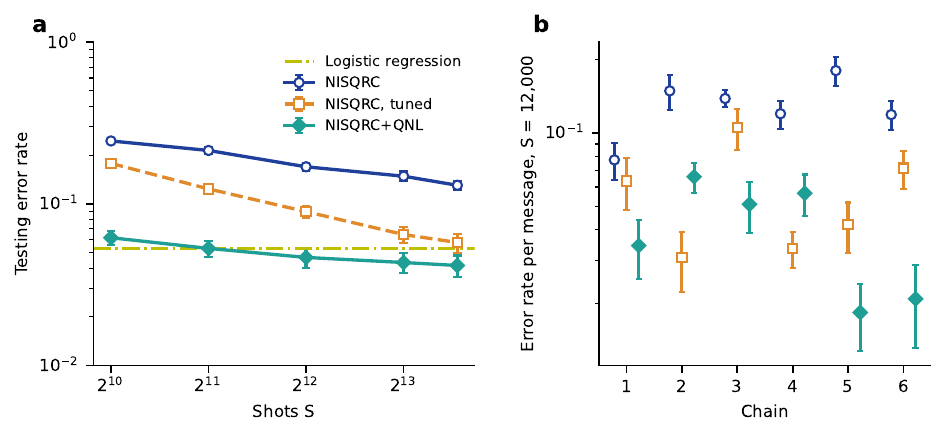}
\caption{\textbf{Frozen configurations on an IBM superconducting processor.} Channel equalization at 20\,dB (100 messages of 20 symbols) on \texttt{ibm\_fez} for the circuit-ansatz NISQRC instance of ref.~\cite{Hu2024} (NISQRC), the same ansatz tuned by random search with 160 candidates (NISQRC, tuned) and the QNL configuration (NISQRC+QNL), all fixed in the same noise-aware simulator before execution. Each circuit runs six disjoint seven-qubit chains, and the method-to-chain assignment rotates between circuits so that every method uses every chain for 16--17 messages. \textbf{a}, Testing error rate versus shots per circuit $S$; error bars, s.d. over 20 random train/test partitions; dash-dotted, logistic regression on $u(n)$. \textbf{b}, Error rate per message at $S=12{,}000$, grouped by chain (mean $\pm$ s.e.m.).}
\label{fig:hardware}
\end{figure*}

\subsection*{A frozen QNL configuration on superconducting hardware}

To examine whether a configuration reorganized in simulation transfers to a real device, we execute the circuit-ansatz NISQRC of ref.~\cite{Hu2024} on the IBM Heron~r2 processor \texttt{ibm\_fez}\cite{IBMQuantum2026}. The reservoir is a seven-qubit chain of alternating readout (R) and memory (M) qubits, R--M--R--M--R--M--R, in which each input is encoded by three Trotter-like layers of single-qubit rotations and nearest-neighbour $ZZ$ interactions, after which the four readout qubits are measured and reset (Methods). QNL is learned in a noise-aware simulator that includes energy relaxation, dephasing and readout errors and omits gate errors and crosstalk. The frozen configuration uses five of the six chain couplers, and each message requires 600 instead of 720 CZ gates at the same depth (Supplementary Table~2).

In a first run with fixed chain assignments, QNL reduces the error rate from $0.097\pm0.009$ to $0.022\pm0.004$ at $S=20{,}000$ (Supplementary Note~8). To separate the effect of QNL from chain-to-chain differences and from the gain that offline tuning alone could provide, a second run adds a tuned baseline obtained by random search in the same simulator with the same budget of 160 candidate configurations, and rotates the assignment of methods to six disjoint chains from circuit to circuit (Fig.~\ref{fig:hardware}). As shown in Fig.~\ref{fig:hardware}a, QNL attains the lowest error rate among the three reservoir configurations at every shot budget: $0.042\pm0.006$ at $S=12{,}000$, against $0.058\pm0.008$ for the tuned baseline and $0.130\pm0.008$ for the reference. Further, QNL falls below the logistic-regression baseline from $S\approx2^{12}$. Paired over the 100 messages, the per-message error rate of QNL is lower than that of the reference (mean difference $0.089$, two-sided Wilcoxon signed-rank test, $P=9\times10^{-16}$) and of the tuned baseline (mean difference $0.016$, $P=0.013$), and QNL has a lower error rate than the reference on all six chains and than the tuned baseline on four (Fig.~\ref{fig:hardware}b). All configurations perform worse on the device than predicted by the noise model, and the ratio of the tuned-baseline to the QNL error rate reduces from $3.8$ in the model to $1.4$ on the device, consistent with the gate errors and crosstalk absent from the model.

\subsection*{One-step-ahead nonlinear forecasting}

Finally, we ask whether plasticity improves prediction of the next output $y(t+1)$ of NARMA-$n$ series from the features at step $t$ (Supplementary Note~9). OLP-weak with QNL attains the lowest mean test root-mean-square error at every order, and its reduction relative to the fixed RSP and OLP-weak reservoirs increases with the order, from 10--13\% for $n=5$ to 17--18\% for $n=10$ and 19--20\% for $n=20$ (Supplementary Fig.~3). With four realizations, the $\pm$s.d. ranges separate at $n=20$ but overlap at $n=5$ and $10$, and these differences are reported as descriptive trends. As in the memory studies, the benefit of plasticity increases with the memory length required by the task.

\section*{Discussion}

Our results show that the physical configuration of a quantum reservoir can be reorganized during learning, by a classical controller that uses evaluations of the task loss, and that this reorganization can improve temporal learning without additional qubits and without changing the measurement protocol or the readout architecture. Plasticity helped most where the task demanded more memory than the fixed reservoir provides: long-term recall, high-order NARMA processing and forecasting, and channel equalization. It redistributes memory, and stronger measurement back-action limits what the reorganized configurations can retain (Fig.~\ref{fig:ew}). This pattern agrees with the motivating analogy, in which transient changes at existing synapses support short-term memory, whereas structural reorganization is associated with long-term retention\cite{Kandel2001,Holtmaat2009}. The analogy concerns design principles: the admission gate of QNL is a supervised, global statistical test on the task loss, and whether local, Hebbian-like rules acting on measurement records could drive the reorganization of quantum reservoirs remains an open question.

The comparison with offline tuning shows what plasticity adds. QNL reached error rates similar to those of a random search over the same number of candidates in simulation (Fig.~\ref{fig:nisqrc}) and remained modestly better on hardware, by a margin that varied across qubit chains (Fig.~\ref{fig:hardware}). The benefit of QNL is that it improves the configuration in use incrementally, by local and interpretable edits, under the back-action and sampling noise of the deployed reservoir, and that it continues to adapt when the task changes (Fig.~\ref{fig:plasticity}), where an offline search would need to be repeated. The measurement budget is important: admission based on shot-sampled losses produced the configurations that were least affected by finite sampling among those compared, in line with the information--disturbance trade-off of monitored reservoirs\cite{MorguiSancho2026} and the role of sampling noise in physical learners\cite{Hu2023}.

The cost of plasticity is paid during learning: the hardware configuration required $\sim10^{9}$ simulated shots, which is why learning was performed in simulation. Sequential tests that stop early, or parallel reservoirs that evaluate candidates simultaneously, could make learning on hardware feasible and allow a deployed reservoir to re-adapt after calibration drift\cite{Klimov2018}. The study has limitations: the reservoirs are small (five to seven qubits), the memory studies use exact expectation values or a Gaussian shot-noise model, the NISQRC and forecasting comparisons rest on eight and four realizations, the hardware evidence consists of two runs on one device, and retention of earlier tasks was not tested. Our results indicate that, for the same quantum reservoir at the same qubit number, plasticity of its physical configuration is a resource for temporal learning that can be added to existing protocols and that can carry over from simulation to a superconducting processor.

\section*{Methods}

\subsection*{Monitored reservoirs and tasks}

The erase-and-write reservoirs follow refs.~\cite{mujal2023,MorguiSancho2026}: five qubits evolve under a transverse-field Ising Hamiltonian, qubit~0 is replaced by the input state $\sqrt{1-s_k}\,|0\rangle+\sqrt{s_k}\,|1\rangle$ at every step, and the 31 features are the expectation values of all non-empty products of Pauli-$Z$ operators. RSP has no back-action, and the online protocols apply non-selective dephasing measurements of strength $\theta=\pi/4$ (OLP-int) or $0.9\pi/2$ (OLP-weak) to every qubit. The STM target at delay $\tau$ is $s_{k-\tau}$, the NARMA-$n$ targets follow refs.~\cite{Atiya2000,fujii2017}, and the capacity for a target is the squared Pearson correlation between the readout output and the target on a held-out segment\cite{Jaeger2002,dambre2012}; the learning loss is one minus the mean capacity across the targets of a task. All models, the amplitude-damping reservoirs and the task definitions are given in Supplementary Notes~2 and~3.

\subsection*{QNL learning rule}

In every block, $K=4$ candidates are nominated. For each candidate, an edit type is drawn with a probability that gives equal weight to a softmax of the eligibility traces and to a uniform choice, so that every edit type remains possible; the edit is applied to a randomly chosen coupler or qubit (or to the evolution time) with a value drawn from a discrete set. For each candidate, we compute a lower confidence bound on its paired loss reduction, using an error level divided among the candidates (and, in the simulations, also among the blocks), and subtract a cost of $\alpha=0.002$ per added coupler. The candidate with the largest resulting margin is admitted if this margin is positive. All eligibility traces then decay by a factor $\lambda=0.8$, and the trace of the admitted edit type receives the margin (Supplementary Note~1 and Supplementary Table~1).

\subsection*{NISQRC, channel equalization and hardware execution}

The NISQRC reservoirs, the channel-equalization task and the readout follow ref.~\cite{Hu2024}. For learning, the seven-qubit chain was simulated exactly in the Hilbert space of its three memory qubits, with relaxation and dephasing of the memory qubits, readout relaxation and assignment errors, and multinomial sampling. On \texttt{ibm\_fez}, six disjoint seven-qubit chains with low CZ and readout errors were selected, and circuits were transpiled with a fixed layout and no routing and executed with the Qiskit Runtime sampler\cite{JavadiAbhari2024}, using mid-circuit measurement and reset\cite{Corcoles2021} and no error mitigation. The error rate was averaged over 20 random partitions of the 100 test messages into training and test halves\cite{Hu2024} (Supplementary Notes~7 and~8).

\subsection*{Statistics}
The independent units are reservoir realizations ($n=\nreal$) for Fig.~\ref{fig:ew}, whole task sequences ($n=\nseq$) for Fig.~\ref{fig:plasticity} and realizations ($n=8$) for Fig.~\ref{fig:nisqrc}. For the hardware runs, error bars give the s.d. over 20 random train/test partitions of the messages, and paired per-message comparisons use two-sided Wilcoxon signed-rank tests\cite{Wilcoxon1945}. Details are given in Supplementary Note~10.

\subsection*{Software}
Simulations used NumPy\cite{Harris2020}, SciPy\cite{Virtanen2020} and scikit-learn\cite{Pedregosa2011}; circuits were built, transpiled and executed with Qiskit and Qiskit Runtime\cite{JavadiAbhari2024}.

\backmatter

\bmhead{Supplementary information}
Supplementary Notes 1--10, Supplementary Figs.~1--3 and Supplementary Tables~1 and~2.

\bmhead{Data availability}
The data supporting the findings of this study are available from the corresponding author upon reasonable request.

\bmhead{Code availability}
The code used in this study is available from the corresponding author upon reasonable request.

\bmhead{Acknowledgments}
We acknowledge the use of IBM Quantum services (open-instance) for this work. The author thanks colleagues at KTH, IBM, and RIKEN for discussions. This work was supported by the European Union under the MULTIPLY-6G
project (No. 101293106) and the QUEST-6G project (No. 101292676). M.R.L. was on sabbatical leave from IBM Research during this work.

\bmhead{Author contributions}
M.R.L. conceived the study, developed the framework, performed the simulations, the hardware experiment and the analysis, and wrote the manuscript.

\bmhead{Competing interests}
The author declares no competing interests.

\bibliography{reference}

@article{Preskill2018,
  author  = {Preskill, John},
  title   = {Quantum computing in the {NISQ} era and beyond},
  journal = {Quantum},
  volume  = {2},
  pages   = {79},
  year    = {2018},
  doi     = {10.22331/q-2018-08-06-79}
}

@article{Biamonte2017,
  author  = {Biamonte, Jacob and Wittek, Peter and Pancotti, Nicola and Rebentrost, Patrick and Wiebe, Nathan and Lloyd, Seth},
  title   = {Quantum machine learning},
  journal = {Nature},
  volume  = {549},
  pages   = {195--202},
  year    = {2017},
  doi     = {10.1038/nature23474}
}

@article{Bharti2022,
  author  = {Bharti, Kishor and Cervera-Lierta, Alba and Kyaw, Thi Ha and Haug, Tobias and Alperin-Lea, Sumner and Anand, Abhinav and Degroote, Matthias and Heimonen, Hermanni and Kottmann, Jakob S. and Menke, Tim and Mok, Wai-Keong and Sim, Sukin and Kwek, Leong-Chuan and Aspuru-Guzik, Al{\'a}n},
  title   = {Noisy intermediate-scale quantum algorithms},
  journal = {Rev. Mod. Phys.},
  volume  = {94},
  pages   = {015004},
  year    = {2022},
  doi     = {10.1103/RevModPhys.94.015004}
}

@article{fujii2017,
  author  = {Fujii, Keisuke and Nakajima, Kohei},
  title   = {Harnessing disordered-ensemble quantum dynamics for machine learning},
  journal = {Phys. Rev. Appl.},
  volume  = {8},
  number  = {2},
  pages   = {024030},
  year    = {2017},
  doi     = {10.1103/PhysRevApplied.8.024030}
}

@article{Nakajima2019,
  author  = {Nakajima, Kohei and Fujii, Keisuke and Negoro, Makoto and Mitarai, Kosuke and Kitagawa, Masahiro},
  title   = {Boosting computational power through spatial multiplexing in quantum reservoir computing},
  journal = {Phys. Rev. Appl.},
  volume  = {11},
  number  = {3},
  pages   = {034021},
  year    = {2019},
  doi     = {10.1103/PhysRevApplied.11.034021}
}

@article{Ghosh2019,
  author  = {Ghosh, Sanjib and Opala, Andrzej and Matuszewski, Micha{\l} and Paterek, Tomasz and Liew, Timothy C. H.},
  title   = {Quantum reservoir processing},
  journal = {npj Quantum Inf.},
  volume  = {5},
  pages   = {35},
  year    = {2019},
  doi     = {10.1038/s41534-019-0149-8}
}

@article{Mujal2021,
  author  = {Mujal, Pere and Mart{\'\i}nez-Pe{\~n}a, Rodrigo and Nokkala, Johannes and Garc{\'\i}a-Beni, Jorge and Giorgi, Gian Luca and Soriano, Miguel C. and Zambrini, Roberta},
  title   = {Opportunities in quantum reservoir computing and extreme learning machines},
  journal = {Adv. Quantum Technol.},
  volume  = {4},
  number  = {8},
  pages   = {2100027},
  year    = {2021},
  doi     = {10.1002/qute.202100027}
}

@article{maass2002,
  author  = {Maass, Wolfgang and Natschl{\"a}ger, Thomas and Markram, Henry},
  title   = {Real-time computing without stable states: a new framework for neural computation based on perturbations},
  journal = {Neural Comput.},
  volume  = {14},
  number  = {11},
  pages   = {2531--2560},
  year    = {2002},
  doi     = {10.1162/089976602760407955}
}

@article{jaeger2004,
  author  = {Jaeger, Herbert and Haas, Harald},
  title   = {Harnessing nonlinearity: predicting chaotic systems and saving energy in wireless communication},
  journal = {Science},
  volume  = {304},
  number  = {5667},
  pages   = {78--80},
  year    = {2004},
  doi     = {10.1126/science.1091277}
}

@article{Lukosevicius2009,
  author  = {Luko{\v s}evi{\v c}ius, Mantas and Jaeger, Herbert},
  title   = {Reservoir computing approaches to recurrent neural network training},
  journal = {Comput. Sci. Rev.},
  volume  = {3},
  number  = {3},
  pages   = {127--149},
  year    = {2009},
  doi     = {10.1016/j.cosrev.2009.03.005}
}

@article{Tanaka2019,
  author  = {Tanaka, Gouhei and Yamane, Toshiyuki and H{\'e}roux, Jean Benoit and Nakane, Ryosho and Kanazawa, Naoki and Takeda, Seiji and Numata, Hidetoshi and Nakano, Daiju and Hirose, Akira},
  title   = {Recent advances in physical reservoir computing: a review},
  journal = {Neural Netw.},
  volume  = {115},
  pages   = {100--123},
  year    = {2019},
  doi     = {10.1016/j.neunet.2019.03.005}
}

@article{dambre2012,
  author  = {Dambre, Joni and Verstraeten, David and Schrauwen, Benjamin and Massar, Serge},
  title   = {Information processing capacity of dynamical systems},
  journal = {Sci. Rep.},
  volume  = {2},
  number  = {1},
  pages   = {514},
  year    = {2012},
  doi     = {10.1038/srep00514}
}

@book{Hebb1949,
  author    = {Hebb, Donald O.},
  title     = {The Organization of Behavior: A Neuropsychological Theory},
  publisher = {Wiley},
  address   = {New York},
  year      = {1949}
}

@article{Bliss1973,
  author  = {Bliss, Timothy V. P. and L{\o}mo, Terje},
  title   = {Long-lasting potentiation of synaptic transmission in the dentate area of the anaesthetized rabbit following stimulation of the perforant path},
  journal = {J. Physiol.},
  volume  = {232},
  number  = {2},
  pages   = {331--356},
  year    = {1973},
  doi     = {10.1113/jphysiol.1973.sp010273}
}

@article{Holtmaat2009,
  author  = {Holtmaat, Anthony and Svoboda, Karel},
  title   = {Experience-dependent structural synaptic plasticity in the mammalian brain},
  journal = {Nat. Rev. Neurosci.},
  volume  = {10},
  number  = {9},
  pages   = {647--658},
  year    = {2009},
  doi     = {10.1038/nrn2699}
}

@article{Xu2009,
  author  = {Xu, Tonghui and Yu, Xinzhu and Perlik, Andrew J. and Tobin, Willie F. and Zweig, Jonathan A. and Tennant, Kelly and Jones, Theresa and Zuo, Yi},
  title   = {Rapid formation and selective stabilization of synapses for enduring motor memories},
  journal = {Nature},
  volume  = {462},
  pages   = {915--919},
  year    = {2009},
  doi     = {10.1038/nature08389}
}

@article{Yang2009,
  author  = {Yang, Guang and Pan, Feng and Gan, Wen-Biao},
  title   = {Stably maintained dendritic spines are associated with lifelong memories},
  journal = {Nature},
  volume  = {462},
  pages   = {920--924},
  year    = {2009},
  doi     = {10.1038/nature08577}
}

@article{lamprecht2004,
  author  = {Lamprecht, Raphael and LeDoux, Joseph},
  title   = {Structural plasticity and memory},
  journal = {Nat. Rev. Neurosci.},
  volume  = {5},
  number  = {1},
  pages   = {45--54},
  year    = {2004},
  doi     = {10.1038/nrn1301}
}

@article{Chklovskii2004,
  author  = {Chklovskii, Dmitri B. and Mel, Bartlett W. and Svoboda, Karel},
  title   = {Cortical rewiring and information storage},
  journal = {Nature},
  volume  = {431},
  pages   = {782--788},
  year    = {2004},
  doi     = {10.1038/nature03012}
}

@article{frey1997,
  author  = {Frey, Uwe and Morris, Richard G. M.},
  title   = {Synaptic tagging and long-term potentiation},
  journal = {Nature},
  volume  = {385},
  number  = {6616},
  pages   = {533--536},
  year    = {1997},
  doi     = {10.1038/385533a0}
}

@article{Gerstner2018,
  author  = {Gerstner, Wulfram and Lehmann, Marco and Liakoni, Vasiliki and Corneil, Dane and Brea, Johanni},
  title   = {Eligibility traces and plasticity on behavioral time scales: experimental support of {neoHebbian} three-factor learning rules},
  journal = {Front. Neural Circuits},
  volume  = {12},
  pages   = {53},
  year    = {2018},
  doi     = {10.3389/fncir.2018.00053}
}

@article{Fremaux2016,
  author  = {Fr{\'e}maux, Nicolas and Gerstner, Wulfram},
  title   = {Neuromodulated spike-timing-dependent plasticity, and theory of three-factor learning rules},
  journal = {Front. Neural Circuits},
  volume  = {9},
  pages   = {85},
  year    = {2016},
  doi     = {10.3389/fncir.2015.00085}
}

@article{fusi2005,
  author  = {Fusi, Stefano and Drew, Patrick J. and Abbott, Larry F.},
  title   = {Cascade models of synaptically stored memories},
  journal = {Neuron},
  volume  = {45},
  number  = {4},
  pages   = {599--611},
  year    = {2005},
  doi     = {10.1016/j.neuron.2005.02.001}
}

@article{benna2016,
  author  = {Benna, Marcus K. and Fusi, Stefano},
  title   = {Computational principles of synaptic memory consolidation},
  journal = {Nat. Neurosci.},
  volume  = {19},
  number  = {12},
  pages   = {1697--1706},
  year    = {2016},
  doi     = {10.1038/nn.4401}
}

@article{Kappel2015,
  author  = {Kappel, David and Habenschuss, Stefan and Legenstein, Robert and Maass, Wolfgang},
  title   = {Network plasticity as {Bayesian} inference},
  journal = {PLoS Comput. Biol.},
  volume  = {11},
  number  = {11},
  pages   = {e1004485},
  year    = {2015},
  doi     = {10.1371/journal.pcbi.1004485}
}

@article{bikic2025,
  author  = {Biki{\'c}, Antonio and Kaspar, Corinna and Pernice, Wolfram H. P.},
  title   = {The cost of unmodeled biological complexity in artificial neural networks},
  journal = {Patterns},
  volume  = {6},
  number  = {10},
  pages   = {101343},
  year    = {2025},
  doi     = {10.1016/j.patter.2025.101343}
}

@article{Stanley2002,
  author  = {Stanley, Kenneth O. and Miikkulainen, Risto},
  title   = {Evolving neural networks through augmenting topologies},
  journal = {Evol. Comput.},
  volume  = {10},
  number  = {2},
  pages   = {99--127},
  year    = {2002},
  doi     = {10.1162/106365602320169811}
}

@article{Mocanu2018,
  author  = {Mocanu, Decebal Constantin and Mocanu, Elena and Stone, Peter and Nguyen, Phuong H. and Gibescu, Madeleine and Liotta, Antonio},
  title   = {Scalable training of artificial neural networks with adaptive sparse connectivity inspired by network science},
  journal = {Nat. Commun.},
  volume  = {9},
  number  = {1},
  pages   = {2383},
  year    = {2018},
  doi     = {10.1038/s41467-018-04316-3}
}

@misc{Bellec2018,
  author       = {Bellec, Guillaume and Kappel, David and Maass, Wolfgang and Legenstein, Robert},
  title        = {Deep rewiring: training very sparse deep networks},
  howpublished = {In \emph{International Conference on Learning Representations (ICLR)}},
  year         = {2018}
}

@article{Bellec2020,
  author  = {Bellec, Guillaume and Scherr, Franz and Subramoney, Anand and Hajek, Elias and Salaj, Darjan and Legenstein, Robert and Maass, Wolfgang},
  title   = {A solution to the learning dilemma for recurrent networks of spiking neurons},
  journal = {Nat. Commun.},
  volume  = {11},
  number  = {1},
  pages   = {3625},
  year    = {2020},
  doi     = {10.1038/s41467-020-17236-y}
}

@article{Kirkpatrick2017,
  author  = {Kirkpatrick, James and Pascanu, Razvan and Rabinowitz, Neil and Veness, Joel and Desjardins, Guillaume and Rusu, Andrei A. and Milan, Kieran and Quan, John and Ramalho, Tiago and Grabska-Barwinska, Agnieszka and Hassabis, Demis and Clopath, Claudia and Kumaran, Dharshan and Hadsell, Raia},
  title   = {Overcoming catastrophic forgetting in neural networks},
  journal = {Proc. Natl Acad. Sci. USA},
  volume  = {114},
  number  = {13},
  pages   = {3521--3526},
  year    = {2017},
  doi     = {10.1073/pnas.1611835114}
}

@article{Zenke2017,
  author  = {Zenke, Friedemann and Poole, Ben and Ganguli, Surya},
  title   = {Continual learning through synaptic intelligence},
  journal = {Proc. Mach. Learn. Res.},
  volume  = {70},
  pages   = {3987--3995},
  year    = {2017}
}

@article{Steil2007,
  author  = {Steil, Jochen J.},
  title   = {Online reservoir adaptation by intrinsic plasticity for backpropagation--decorrelation and echo state learning},
  journal = {Neural Netw.},
  volume  = {20},
  number  = {3},
  pages   = {353--364},
  year    = {2007},
  doi     = {10.1016/j.neunet.2007.04.011}
}

@article{Schrauwen2008,
  author  = {Schrauwen, Benjamin and Wardermann, Marion and Verstraeten, David and Steil, Jochen J. and Stroobandt, Dirk},
  title   = {Improving reservoirs using intrinsic plasticity},
  journal = {Neurocomputing},
  volume  = {71},
  number  = {7--9},
  pages   = {1159--1171},
  year    = {2008},
  doi     = {10.1016/j.neucom.2007.12.020}
}

@article{Lazar2009,
  author  = {Lazar, Andreea and Pipa, Gordon and Triesch, Jochen},
  title   = {{SORN}: a self-organizing recurrent neural network},
  journal = {Front. Comput. Neurosci.},
  volume  = {3},
  pages   = {23},
  year    = {2009},
  doi     = {10.3389/neuro.10.023.2009}
}

@article{Dohare2024,
  author  = {Dohare, Shibhansh and Hernandez-Garcia, J. Fernando and Lan, Qingfeng and Rahman, Parash and Mahmood, A. Rupam and Sutton, Richard S.},
  title   = {Loss of plasticity in deep continual learning},
  journal = {Nature},
  volume  = {632},
  number  = {8026},
  pages   = {768--774},
  year    = {2024},
  doi     = {10.1038/s41586-024-07711-7}
}

@article{mujal2023,
  author  = {Mujal, Pere and Mart{\'\i}nez-Pe{\~n}a, Rodrigo and Giorgi, Gian Luca and Soriano, Miguel C. and Zambrini, Roberta},
  title   = {Time-series quantum reservoir computing with weak and projective measurements},
  journal = {npj Quantum Inf.},
  volume  = {9},
  number  = {1},
  pages   = {16},
  year    = {2023},
  doi     = {10.1038/s41534-023-00682-z}
}

@misc{MorguiSancho2026,
  author       = {Morgu{\'\i}-Sancho, Oriol and Manzano, Gonzalo and Giorgi, Gian Luca and Zambrini, Roberta},
  title        = {General theory of monitored quantum reservoir computing},
  howpublished = {Preprint at https://arxiv.org/abs/2607.14311},
  year         = {2026}
}

@article{Hu2024,
  author  = {Hu, Fangjun and Khan, Saeed A. and Bronn, Nicholas T. and Angelatos, Gerasimos and Rowlands, Graham E. and Ribeill, Guilhem J. and T{\"u}reci, Hakan E.},
  title   = {Overcoming the coherence time barrier in quantum machine learning on temporal data},
  journal = {Nat. Commun.},
  volume  = {15},
  pages   = {7491},
  year    = {2024},
  doi     = {10.1038/s41467-024-51162-7}
}

@article{Hu2023,
  author  = {Hu, Fangjun and Angelatos, Gerasimos and Khan, Saeed A. and Vives, Marti and T{\"u}reci, Esin and Bello, Leon and Rowlands, Graham E. and Ribeill, Guilhem J. and T{\"u}reci, Hakan E.},
  title   = {Tackling sampling noise in physical systems for machine learning applications: fundamental limits and eigentasks},
  journal = {Phys. Rev. X},
  volume  = {13},
  pages   = {041020},
  year    = {2023},
  doi     = {10.1103/PhysRevX.13.041020}
}

@article{Cerezo2021,
  author  = {Cerezo, M. and Arrasmith, Andrew and Babbush, Ryan and Benjamin, Simon C. and Endo, Suguru and Fujii, Keisuke and McClean, Jarrod R. and Mitarai, Kosuke and Yuan, Xiao and Cincio, Lukasz and Coles, Patrick J.},
  title   = {Variational quantum algorithms},
  journal = {Nat. Rev. Phys.},
  volume  = {3},
  number  = {9},
  pages   = {625--644},
  year    = {2021},
  doi     = {10.1038/s42254-021-00348-9}
}

@article{McClean2018,
  author  = {McClean, Jarrod R. and Boixo, Sergio and Smelyanskiy, Vadim N. and Babbush, Ryan and Neven, Hartmut},
  title   = {Barren plateaus in quantum neural network training landscapes},
  journal = {Nat. Commun.},
  volume  = {9},
  number  = {1},
  pages   = {4812},
  year    = {2018},
  doi     = {10.1038/s41467-018-07090-4}
}

@article{Larocca2025,
  author  = {Larocca, Mart{\'\i}n and Thanasilp, Supanut and Wang, Samson and Sharma, Kunal and Biamonte, Jacob and Coles, Patrick J. and Cincio, Lukasz and McClean, Jarrod R. and Holmes, Zo{\"e} and Cerezo, M.},
  title   = {Barren plateaus in variational quantum computing},
  journal = {Nat. Rev. Phys.},
  volume  = {7},
  pages   = {174--189},
  year    = {2025},
  doi     = {10.1038/s42254-025-00813-9}
}

@article{Grimsley2019,
  author  = {Grimsley, Harper R. and Economou, Sophia E. and Barnes, Edwin and Mayhall, Nicholas J.},
  title   = {An adaptive variational algorithm for exact molecular simulations on a quantum computer},
  journal = {Nat. Commun.},
  volume  = {10},
  number  = {1},
  pages   = {3007},
  year    = {2019},
  doi     = {10.1038/s41467-019-10988-2}
}

@article{Du2022,
  author  = {Du, Yuxuan and Huang, Tao and You, Shan and Hsieh, Min-Hsiu and Tao, Dacheng},
  title   = {Quantum circuit architecture search for variational quantum algorithms},
  journal = {npj Quantum Inf.},
  volume  = {8},
  number  = {1},
  pages   = {62},
  year    = {2022},
  doi     = {10.1038/s41534-022-00570-y}
}

@article{Jiang2022,
  author  = {Jiang, Wenjie and Lu, Zhide and Deng, Dong-Ling},
  title   = {Quantum continual learning overcoming catastrophic forgetting},
  journal = {Chin. Phys. Lett.},
  volume  = {39},
  number  = {5},
  pages   = {050303},
  year    = {2022},
  doi     = {10.1088/0256-307X/39/5/050303}
}

@article{Chen2026,
  author  = {Chen, Yu-Qin and Zhang, Shi-Xin},
  title   = {Intrinsic preservation of plasticity in continual quantum learning},
  journal = {PRX Quantum},
  volume  = {7},
  number  = {3},
  pages   = {033003},
  year    = {2026},
  doi     = {10.1103/ry52-85ll}
}

@article{Bergstra2012,
  author  = {Bergstra, James and Bengio, Yoshua},
  title   = {Random search for hyper-parameter optimization},
  journal = {J. Mach. Learn. Res.},
  volume  = {13},
  pages   = {281--305},
  year    = {2012}
}

@article{Atiya2000,
  author  = {Atiya, Amir F. and Parlos, Alexander G.},
  title   = {New results on recurrent network training: unifying the algorithms and accelerating convergence},
  journal = {IEEE Trans. Neural Netw.},
  volume  = {11},
  number  = {3},
  pages   = {697--709},
  year    = {2000},
  doi     = {10.1109/72.846741}
}

@misc{IBMQuantum2026,
  author       = {{IBM Quantum}},
  title        = {{IBM Quantum Platform}: processor ibm\_fez ({Heron} r2)},
  howpublished = {https://quantum.cloud.ibm.com},
  year         = {2026}
}

@article{Klimov2018,
  author  = {Klimov, P. V. and Kelly, J. and Chen, Z. and Neeley, M. and Megrant, A. and others},
  title   = {Fluctuations of energy-relaxation times in superconducting qubits},
  journal = {Phys. Rev. Lett.},
  volume  = {121},
  number  = {9},
  pages   = {090502},
  year    = {2018},
  doi     = {10.1103/PhysRevLett.121.090502}
}

@article{Pedregosa2011,
  author  = {Pedregosa, Fabian and Varoquaux, Ga{\"e}l and Gramfort, Alexandre and Michel, Vincent and Thirion, Bertrand and Grisel, Olivier and Blondel, Mathieu and Prettenhofer, Peter and Weiss, Ron and Dubourg, Vincent and Vanderplas, Jake and Passos, Alexandre and Cournapeau, David and Brucher, Matthieu and Perrot, Matthieu and Duchesnay, {\'E}douard},
  title   = {Scikit-learn: machine learning in {Python}},
  journal = {J. Mach. Learn. Res.},
  volume  = {12},
  pages   = {2825--2830},
  year    = {2011}
}

@misc{JavadiAbhari2024,
  author       = {Javadi-Abhari, Ali and others},
  title        = {Quantum computing with {Qiskit}},
  howpublished = {Preprint at https://arxiv.org/abs/2405.08810},
  year         = {2024}
}

@article{Corcoles2021,
  author  = {C{\'o}rcoles, A. D. and Takita, Maika and Inoue, Ken and Lekuch, Scott and Minev, Zlatko K. and Chow, Jerry M. and Gambetta, Jay M.},
  title   = {Exploiting dynamic quantum circuits in a quantum algorithm with superconducting qubits},
  journal = {Phys. Rev. Lett.},
  volume  = {127},
  number  = {10},
  pages   = {100501},
  year    = {2021},
  doi     = {10.1103/PhysRevLett.127.100501}
}

@article{Wilcoxon1945,
  author  = {Wilcoxon, Frank},
  title   = {Individual comparisons by ranking methods},
  journal = {Biometrics Bull.},
  volume  = {1},
  number  = {6},
  pages   = {80--83},
  year    = {1945},
  doi     = {10.2307/3001968}
}

@article{Harris2020,
  author  = {Harris, Charles R. and others},
  title   = {Array programming with {NumPy}},
  journal = {Nature},
  volume  = {585},
  number  = {7825},
  pages   = {357--362},
  year    = {2020},
  doi     = {10.1038/s41586-020-2649-2}
}

@article{Virtanen2020,
  author  = {Virtanen, Pauli and others},
  title   = {{SciPy} 1.0: fundamental algorithms for scientific computing in {Python}},
  journal = {Nat. Methods},
  volume  = {17},
  number  = {3},
  pages   = {261--272},
  year    = {2020},
  doi     = {10.1038/s41592-019-0686-2}
}

@article{Kandel2001,
  author  = {Kandel, Eric R.},
  title   = {The molecular biology of memory storage: a dialogue between genes and synapses},
  journal = {Science},
  volume  = {294},
  number  = {5544},
  pages   = {1030--1038},
  year    = {2001},
  doi     = {10.1126/science.1067020}
}

@techreport{Jaeger2002,
  author      = {Jaeger, Herbert},
  title       = {Short term memory in echo state networks},
  institution = {GMD -- German National Research Center for Information Technology},
  type        = {GMD Report},
  number      = {152},
  year        = {2002}
}

@book{Wiseman2009,
  author    = {Wiseman, Howard M. and Milburn, Gerard J.},
  title     = {Quantum Measurement and Control},
  publisher = {Cambridge University Press},
  address   = {Cambridge},
  year      = {2009},
  doi       = {10.1017/CBO9780511813948}
}

@book{Nielsen2010,
  author    = {Nielsen, Michael A. and Chuang, Isaac L.},
  title     = {Quantum Computation and Quantum Information},
  edition   = {10th anniversary},
  publisher = {Cambridge University Press},
  address   = {Cambridge},
  year      = {2010},
  doi       = {10.1017/CBO9780511976667}
}

@techreport{jaeger2001echo,
  author      = {Jaeger, Herbert},
  title       = {The ``echo state'' approach to analysing and training recurrent neural networks},
  institution = {GMD -- German National Research Center for Information Technology},
  address     = {Bonn, Germany},
  type        = {GMD Report},
  number      = {148},
  year        = {2001}
}

@book{SuttonBarto2018,
  author    = {Sutton, Richard S. and Barto, Andrew G.},
  title     = {Reinforcement Learning: An Introduction},
  edition   = {2nd},
  publisher = {MIT Press},
  address   = {Cambridge, MA},
  year      = {2018}
}

@article{Dunn1961,
  author  = {Dunn, Olive Jean},
  title   = {Multiple comparisons among means},
  journal = {J. Am. Stat. Assoc.},
  volume  = {56},
  number  = {293},
  pages   = {52--64},
  year    = {1961},
  doi     = {10.1080/01621459.1961.10482090}
}

@article{Student1908,
  author  = {{Student}},
  title   = {The probable error of a mean},
  journal = {Biometrika},
  volume  = {6},
  number  = {1},
  pages   = {1--25},
  year    = {1908},
  doi     = {10.2307/2331554}
}

\clearpage
\onecolumn
\setcounter{equation}{0}\renewcommand{\theequation}{S\arabic{equation}}
\setcounter{figure}{0}\renewcommand{\figurename}{Supplementary Fig.}
\setcounter{table}{0}\renewcommand{\tablename}{Supplementary Table}
\section*{Supplementary Information}
\section*{Supplementary Note 1: QNL learning rule}
Each plasticity block implements equation~(2) of the main text as follows. We generate a candidate by drawing an edit type $\kappa\in\mathcal K$ (for example, ``add a coupler'' or ``change the clock'') with probability $p_\kappa$ and applying it to a uniformly chosen coupler or qubit (or, for ``change the clock'', to the evolution time), with a value drawn uniformly from the alphabet listed in Supplementary Table~\ref{stab:settings}. The edit ``remove'' keeps at least one coupler (three on the hardware chain), and ``add'' activates an inactive coupler. At the level of the reservoir couplers, these add and remove edits are analogous to the formation and elimination of synaptic connections in structural plasticity~\cite{Holtmaat2009,Chklovskii2004}, and the rewiring of connections in models of network plasticity~\cite{Kappel2015} and in sparse artificial networks~\cite{Bellec2018,Mocanu2018}. Since the base reservoirs of the erase-and-write and amplitude-damping studies are fully connected, ``add'' can restore a coupler removed earlier; the monitoring strength is not edited in these studies. QNL preserves the number of qubits, the functional form of the input encoding, the measurement protocol and the type of readout. In the NISQRC models, the edits include the input-field coefficients or the input-dependent gate angles (Supplementary Table~\ref{stab:settings}), and the readout weights are refitted during every loss evaluation.

In each block, we evaluate the current configuration $\boldsymbol\theta$ and the $K$ candidates $\boldsymbol\theta_c$, $c=1,\dots,K$, on the same $\nstr$ freshly drawn streams $\xi_1,\dots,\xi_{\nstr}$, which gives the paired loss reductions $g_{c,r}=L(\boldsymbol\theta;\xi_r)-L(\boldsymbol\theta_c;\xi_r)$ with mean $\bar g_c$ and sample standard deviation $s_c$. Evaluating the loss difference within each shared stream reduces the variability caused by differences in stream difficulty. The lower confidence bound on the improvement of candidate $c$ is
\begin{equation}
\mathrm{LCB}_c=\bar g_c-t_{1-\delta',\,\nstr-1}\,\frac{s_c}{\sqrt{\nstr}},
\label{eq:lcb}
\end{equation}
where $t_{1-\delta',\,\nstr-1}$ is the Student-$t$ quantile~\cite{Student1908} at a Bonferroni-corrected one-sided level $\delta'$\cite{Dunn1961}. The admission margin of equation~(2) of the main text is
\begin{equation}
\Gamma\big(\boldsymbol\theta_c;\boldsymbol\theta\big)=\mathrm{LCB}_c-\alpha\,[\Delta N_c]_+,
\label{eq:admit}
\end{equation}
where $\Delta N_c=N(\boldsymbol\theta_c)-N(\boldsymbol\theta)$, $N(\boldsymbol\theta)$ is the number of active couplers and $[z]_+=\max(z,0)$; the cost $\alpha$ applies to added couplers, and removing a coupler has no cost. This term acts as a cost on connectivity, motivated by the limited number of synaptic connections in models of cortical rewiring~\cite{Chklovskii2004}. The maximizing candidate $c^\ast$ is admitted if $\Gamma(\boldsymbol\theta_{c^\ast};\boldsymbol\theta)>0$; otherwise, we retain the current configuration. Each edit type carries an eligibility trace $e_\kappa$, which decays by a factor $\lambda$ in every block and receives the positive margin when an edit of that type is admitted. The edit types of the $K$ candidates in equation~(2) of the main text are then drawn with probabilities $p_\kappa$:
\begin{align}
e_\kappa&\leftarrow\lambda\,e_\kappa+\big[\Gamma(\boldsymbol\theta_{c^\ast};\boldsymbol\theta)\big]_+\,\mathbf 1_{\{\kappa=\kappa(c^\ast)\}},\label{eq:elig}\\
p_\kappa&=\frac12\,\frac{e^{e_\kappa/T}}{\sum_{\kappa'\in\mathcal K}e^{e_{\kappa'}/T}}+\frac{1}{2|\mathcal K|},\label{eq:nominate}
\end{align}
where $\mathbf 1_{\{\cdot\}}$ is the indicator function. If no candidate is admitted, equation~(\ref{eq:elig}) reduces to $e_\kappa\leftarrow\lambda e_\kappa$ for every edit type. The probabilities $p_\kappa$ sum to one, and the uniform floor $p_\kappa\geq1/(2|\mathcal K|)$ keeps every edit type possible. Equation~(\ref{eq:elig}) is an accumulating eligibility trace, and equation~(\ref{eq:nominate}) is a Boltzmann (softmax) selection rule mixed with uniform exploration; both rules are standard in reinforcement learning~\cite{SuttonBarto2018}. The form of equation~(\ref{eq:elig}) is inspired by synaptic tagging~\cite{frey1997} and by three-factor learning rules~\cite{Fremaux2016,Gerstner2018}, in which a decaying eligibility trace marks a recent change and a later global signal determines whether it is consolidated; in QNL, the admission margin $\Gamma$ acts as this global signal. The analogy refers to the structure of the update. In QNL, these rules act on edit types. We use $K=4$, $\lambda=0.8$, $T=0.05$ and $\alpha=0.002$ in all experiments.

The error level $\delta'$ of equation~(\ref{eq:lcb}) is Bonferroni-corrected over the $K$ candidates. In the simulations of Figs.~2 and~3 and Supplementary Figs.~\ref{sfig:ad}--\ref{sfig:forecast}, the correction extends to the $B$ blocks of a learning run, and $\delta'=0.05/(KB)$ allocates a nominal family-wise error level of 0.05 over the candidates and blocks of each learning run (or each task-adaptation run). For the NISQRC studies, we use $\delta'=0.3/K$, which allocates a nominal level of 0.3 within each block and does not control the error level over the whole run. The bound is nominal: it assumes that the paired gains across independently drawn streams are approximately normal.

To quantify the ability to keep learning after a task change, in the spirit of loss-of-plasticity analyses~\cite{Dohare2024}, we define the paired plasticity on task $k$ as
\begin{equation}
\Pi_k=\frac{L_k^{\rm before}-L_k^{\rm after}}{L_k^{\rm before}},
\label{eq:plasticity}
\end{equation}
where $L_k^{\rm before}$ and $L_k^{\rm after}$ are the losses on task $k$ of the configuration inherited from task $k-1$ (the base reservoir for $k=1$) and of the adapted configuration, respectively, evaluated on the same held-out streams. Equation~(\ref{eq:plasticity}) is defined for $L_k^{\rm before}>0$; when $L_k^{\rm before}=0$, the absolute loss change $L_k^{\rm before}-L_k^{\rm after}$ can be used instead. For a fixed reservoir evaluated with the same streams and readout-fitting procedure, $L_k^{\rm before}=L_k^{\rm after}$ and $\Pi_k=0$.

\begin{table}[t]
\centering\sffamily\footnotesize
\caption{\textbf{QNL settings.} $K$, candidates per block; $\nstr$, paired streams per block; $B$, blocks per learning run (for task switches, blocks after each switch); $\delta'$, one-sided error level of the lower confidence bound; $\alpha$, cost per added coupler. All experiments use eligibility decay $\lambda=0.8$ and temperature $T=0.05$ (equations~(\ref{eq:elig}) and~(\ref{eq:nominate})).}
\label{stab:settings}
\renewcommand{\arraystretch}{1.2}\setlength{\tabcolsep}{3pt}
\begin{tabular}{@{}>{\raggedright\arraybackslash}p{25mm}>{\raggedright\arraybackslash}p{55mm}ccccc>{\raggedright\arraybackslash}p{25mm}@{}}
\toprule
Experiment & Editable configuration (alphabet) & $K$ & $\nstr$ & $B$ & $\delta'$ & $\alpha$ & Learning loss and features \\
\midrule
Erase-and-write and amplitude damping (Fig.~2, Supplementary Figs.~\ref{sfig:ad} and~\ref{sfig:shots}) & couplers: add, remove, change ($J_{ij}\in\{\pm0.3,\pm0.6,\pm1\}$); fields $h_i\in\{0,0.5,1,1.5,2\}$; clock $\Delta t\in\{0.5,1,1.5,2,3\}$ & 4 & 16 & 30 & $0.05/(KB)$ & 0.002 & $1-\overline C$; exact features \\
Task switches (Fig.~3) & as above & 4 & 16 & 12 per task & $0.05/(KB)$ & 0.002 & $1-\overline C$ of the current task; exact features \\
Forecasting (Supplementary Fig.~\ref{sfig:forecast}) & as above & 4 & 8 & 10 & $0.05/(KB)$ & 0.002 & forecast NMSE, $y(t+1)$; exact features \\
NISQRC, Hamiltonian ansatz (Fig.~4) & couplers: add, remove, change ($J_{ij}\in\{0.1,0.25,0.5,0.75,1\}$); $\eta^x_i\in\{0.5,1,2,3,4\}$; $\eta^z_i\in\{0.25,0.5,0.75,1,1.5\}$; $\Delta t\in\{0.5,1,1.5,2\}$ & 4 & 4 & 20 & $0.3/K$ & 0.002 & NMSE surrogate; $S=10^5$ shots, 20\,dB \\
NISQRC, circuit ansatz (Fig.~5) & couplers: add, remove ($\geq3$ kept), change ($J\in\{0.5,0.75,1,1.25,1.5\}$); $\theta^x_i\in\{0.2,\tfrac13,0.5,\tfrac23,0.9\}$; $\theta^z_i\in\{0.1,\tfrac16,0.25,\tfrac13,0.5\}$; $\theta^I_i\in\{0.02,\tfrac1{30},0.05,\tfrac1{15},0.1\}$ & 4 & 4 & 40 & $0.3/K$ & 0.002 & NMSE surrogate; $S=2^{15}$ shots, noise model, 20\,dB \\
\bottomrule
\end{tabular}
\end{table}

\section*{Supplementary Note 2: Monitored reservoir models}
\paragraph{Erase-and-write reservoirs.} For the reservoirs of Figs.~2 and~3 and Supplementary Figs.~\ref{sfig:shots} and~\ref{sfig:forecast}, we follow refs.~\cite{mujal2023,MorguiSancho2026}. We set $\hbar=1$ and use dimensionless units: Hamiltonian coefficients are expressed in units of an arbitrary energy scale and evolution times in units of its inverse. Five qubits evolve under the transverse-field Ising Hamiltonian
\begin{equation}
H(\boldsymbol\theta)=\sum_{i>j}J_{ij}X_iX_j+\sum_{i}h_iZ_i
\label{eq:tfim}
\end{equation}
for a time $\Delta t$ per input, $U(\boldsymbol\theta)=e^{-iH(\boldsymbol\theta)\Delta t}$, with configuration $\boldsymbol\theta=(\{J_{ij}\},\{h_i\},\Delta t)$. For the base reservoirs, we draw $J_{ij}\sim U[-1,1]$ independently for all ten pairs in every realization, and set $h_i=1$ and $\Delta t=2$. At step $k$, qubit~0 is replaced by $|\psi(s_k)\rangle=\sqrt{1-s_k}\,|0\rangle+\sqrt{s_k}\,|1\rangle$, and the reservoir evolves as in refs.~\cite{mujal2023,MorguiSancho2026},
\begin{equation}
\rho_k=U(\boldsymbol\theta)\big[\psi_k\otimes\mathrm{Tr}_0\,\tilde\rho_{k-1}\big]U(\boldsymbol\theta)^\dagger,
\label{eq:ew}
\end{equation}
where $\psi_k=|\psi(s_k)\rangle\langle\psi(s_k)|$ and $\tilde\rho_{k-1}$ is the state after the measurement of step $k-1$. We use as features the 31 expectation values $x_{k,S}=\mathrm{Tr}(\rho_kZ_S)$, $Z_S=\prod_{i\in S}Z_i$, for all non-empty subsets $S$ of the five qubits. In the online protocols, every qubit is monitored by a non-selective dephasing measurement of strength $\theta$, which can be implemented through a probe qubit~\cite{Wiseman2009,MorguiSancho2026} and maps $\langle a|\rho_k|b\rangle\mapsto\langle a|\rho_k|b\rangle\prod_{i:\,a_i\neq b_i}\sin\theta_i$ in the computational basis to give $\tilde\rho_k$; OLP-int uses $\theta=\pi/4$ and OLP-weak uses $\theta=0.9\pi/2$. In RSP, the reservoir is re-prepared and the input history is replayed before every measurement, giving $\tilde\rho_k=\rho_k$. The features are computed from the pre-measurement state $\rho_k$; in Figs.~2 and~3 and Supplementary Figs.~\ref{sfig:ad} and~\ref{sfig:forecast}, they are exact expectation values.

\paragraph{Amplitude-damping-monitored reservoirs.} In Supplementary Fig.~\ref{sfig:ad}, we encode the input by a rotation of qubit~0, $\rho\mapsto R_k\rho R_k^\dagger$ with $R_k=\sqrt{1-s_k}\,I-i\sqrt{s_k}\,Y_0$ ($I$, identity), which maps $|0\rangle$ to $|\psi(s_k)\rangle$. The reservoir then evolves under the unitary $U(\boldsymbol\theta)=e^{-iH(\boldsymbol\theta)\Delta t}$ generated by equation~(\ref{eq:tfim}), the features are read out, and every qubit undergoes the non-selective amplitude-damping measurement of ref.~\cite{MorguiSancho2026}, with Kraus operators~\cite{Nielsen2010} $\Omega_0=|0\rangle\langle0|+\cos(\theta/2)|1\rangle\langle1|$ and $\Omega_1=\sin(\theta/2)|0\rangle\langle1|$. OLP-int uses $\theta=0.4\pi$ (decay probability $0.35$) and OLP-weak uses $\theta=0.2\pi$ (decay probability $0.095$). The two monitoring schemes use opposite conventions: for amplitude damping, a larger $\theta$ gives stronger monitoring; for dephasing, $\theta=\pi/2$ gives no back-action.

\section*{Supplementary Note 3: Tasks, readout and capacities}
We draw the inputs as $s_k\sim U[0,1]$ for the memory tasks and $s_k\sim U[0,0.2]$ for the NARMA tasks. The STM target at delay $\tau$ is $y_k=s_{k-\tau}$, and the NARMA-$n$ target follows~\cite{Atiya2000,fujii2017,MorguiSancho2026}
\begin{equation}
\begin{aligned}
y_k={}&0.3\,y_{k-1}+0.05\,y_{k-1}\sum_{i=1}^{n}y_{k-i}\\
&+1.5\,s_{k-n}\,s_{k-1}+0.1.
\end{aligned}
\label{eq:narma}
\end{equation}
We fit the linear readout $\hat y_k=\mathbf w^{\!\top}[\mathbf x_k;1]$ by least squares (Moore--Penrose pseudo-inverse) on the training segment. Following refs.~\cite{Jaeger2002,dambre2012}, we define the capacity for a target as the squared Pearson correlation between $\hat y_k$ and $y_k$ on the test segment, $C=\mathrm{cov}^2(\hat y,y)/[\mathrm{var}(\hat y)\,\mathrm{var}(y)]\in[0,1]$. This expression assumes nonzero target and prediction variances. Each evaluation stream has 3,000 steps (1,000 washout, 1,000 training and 1,000 test). We report $\sum_{\tau=0}^{20}C_\tau$ (STM), $\sum_{\tau=15}^{30}C_\tau$ (LTM) and the mean of $C_n$ over $n=2,\dots,10$ (NARMA). For a target set $\mathcal A$, we define the learning loss as $L=1-\overline C$, where $\overline C=|\mathcal A|^{-1}\sum_{a\in\mathcal A}C_a$ is the mean capacity across the targets of the task, computed on learning streams of 1,100 steps (200 washout, 500 training and 400 validation). The learning and evaluation streams are drawn independently.

\section*{Supplementary Note 4: Plasticity under amplitude-damping monitoring}
To test whether the results of Fig.~2 depend on the reservoir map, we apply QNL to monitored unitary reservoirs in which the input is encoded by a rotation and the memory fades through amplitude-damping measurements of every qubit~\cite{MorguiSancho2026}, a non-unital monitoring scheme that provides the dissipation required for fading memory (Supplementary Note~2). In Supplementary Fig.~\ref{sfig:ad}, QNL again increases the summed STM capacity, by $0.99$ for OLP-int ($\theta=0.4\pi$; from $5.71$ to $6.70$) and by $0.43$ for OLP-weak ($\theta=0.2\pi$; from $6.22$ to $6.65$), mainly by extending the capacity at intermediate delays ($\tau\approx5$--$9$). These gains are smaller than those obtained with the erase-and-write map, which points to a dependence on the reservoir map; for this map, the memory profile appears to be dominated by the amplitude-damping monitoring, which QNL does not edit.

\begin{figure}[t]
\centering
\includegraphics[width=0.92\linewidth]{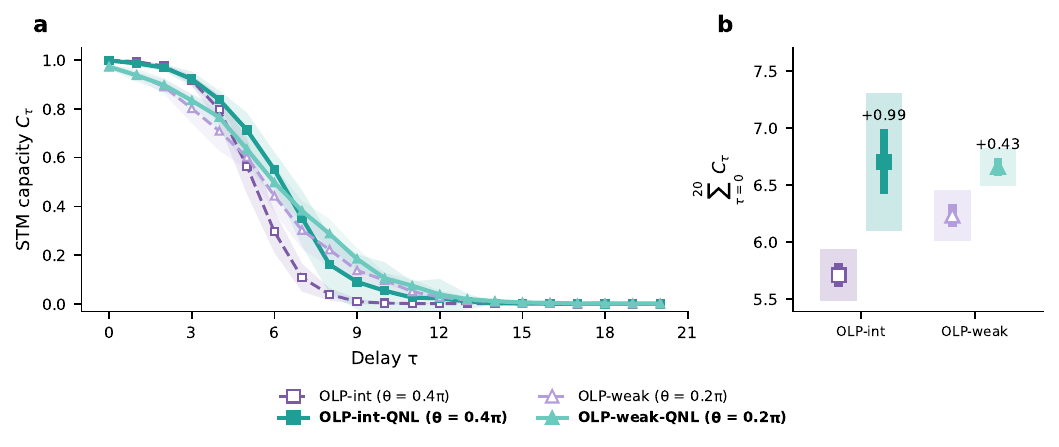}
\caption{\textbf{Plasticity under amplitude-damping monitoring.} Monitored unitary reservoirs~\cite{MorguiSancho2026} in which the input rotates qubit~0 and every qubit is monitored by an amplitude-damping measurement of strength $\theta$ (decay probability $\sin^2(\theta/2)$ per step); OLP-int, $\theta=0.4\pi$; OLP-weak, $\theta=0.2\pi$. Purple, fixed reservoirs; teal, after QNL trained for STM. \textbf{a}, STM capacity $C_\tau$. \textbf{b}, $\sum_{\tau=0}^{20}C_\tau$ with paired mean gains. Lines, bands, boxes and bars as in Fig.~2; $n=\nreal$ reservoir realizations.}
\label{sfig:ad}
\end{figure}

\section*{Supplementary Note 5: Finite measurement budgets}
\begin{figure}[t]
\centering
\includegraphics[width=0.92\linewidth]{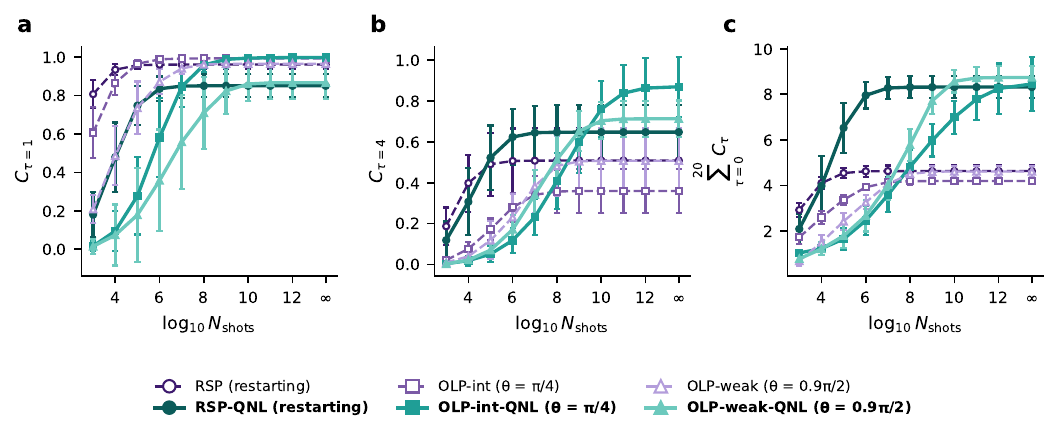}
\caption{\textbf{Effect of finite measurement ensembles.} The STM reservoirs of Fig.~2a,b re-evaluated with 15 features ($\langle Z_i\rangle$, $\langle Z_iZ_j\rangle$) estimated from $N_{\rm shots}$ measurements per time step, modelled as independent Gaussian errors of s.d. $\sigma_S/\sqrt{N_{\rm shots}}$, where $\sigma_S=\prod_{i\in S}(\cos\theta_i)^{-1}$ under back-action and $\sigma_S=1$ for RSP (Supplementary Note~5); $\infty$, exact expectation values. \textbf{a}, $C_{\tau=1}$. \textbf{b}, $C_{\tau=4}$. \textbf{c}, $\sum_{\tau=0}^{20}C_\tau$. Points, mean over $n=\nreal$ reservoir realizations; error bars, s.d. The horizontal axis counts measurements per time step and does not equalize experimental cost: RSP replays the input history for every measurement. QNL was trained with exact features.}
\label{sfig:shots}
\end{figure}

The capacities in Fig.~2 are computed from exact expectation values. On hardware, the features are estimated from a finite number of measurements, and weak monitoring extracts little information per measurement: an expectation value estimated from dephasing measurements of strength $\theta$ carries an error amplified by $1/\cos\theta$ for every qubit in its support~\cite{MorguiSancho2026}. For this reason, we re-evaluate the reservoirs learned for STM with the 15 features $\langle Z_i\rangle$ and $\langle Z_iZ_j\rangle$ estimated from $N_{\rm shots}$ measurements per time step (Supplementary Fig.~\ref{sfig:shots}).

Following ref.~\cite{MorguiSancho2026}, an expectation value estimated from $N_{\rm shots}$ outcomes of dephasing measurements of strength $\theta$ requires a rescaling by $\prod_{i\in S}(\cos\theta_i)^{-1}$, which amplifies its statistical error. We model this effect by adding independent Gaussian noise of s.d. $\sigma_S/\sqrt{N_{\rm shots}}$ to each feature, with $\sigma_S=\prod_{i\in S}(\cos\theta_i)^{-1}$ for the online protocols and $\sigma_S=1$ for RSP, the maximum single-shot standard deviation of a $\pm1$ observable. We average the noise replicates within each reservoir realization and use the same standard-normal draws for all methods. This model treats the features independently and does not simulate measurement trajectories.

The benefit of reorganization depends on the measurement budget (Supplementary Fig.~\ref{sfig:shots}). With RSP, the learned reservoir is close to the fixed one at $10^4$ measurements and reaches a higher capacity from $10^5$ ($\sum_{\tau=0}^{20}C_\tau\approx6.5$ versus $\approx4.6$), saturating near $8.3$ against $4.6$ above $\sim10^7$ (Supplementary Fig.~\ref{sfig:shots}c). With online monitoring, the learned reservoirs perform worse than the fixed ones up to $\sim10^7$ measurements per step and better from $10^8$, eventually reaching $8.5$--$8.8$ against $4.2$--$4.6$. The single-delay capacities (Supplementary Fig.~\ref{sfig:shots}a,b) help explain this behaviour: QNL moves memory from the shortest delays, which are carried by strong feature components, to longer delays carried by weaker components that require larger ensembles to resolve. With these 15 features, QNL lowers $C_{\tau=1}$ of the RSP and OLP-weak reservoirs to about $0.85$ even with exact expectation values (with all 31 features, $C_1\approx0.91$; Fig.~2a). This trade-off is likely related to training with exact features. For deployment, the loss used for admission is best estimated at the intended measurement budget, as in the NISQRC and hardware experiments of the main text.

\section*{Supplementary Note 6: Task-switch protocol}
In each sequence (Fig.~3), we draw 16 tasks uniformly from the pool of four STM windows and three NARMA orders, excluding immediate repetitions, and start from a random base reservoir shared by the three protocols. After each switch, QNL runs for the number of blocks listed in Supplementary Table~\ref{stab:settings}, starting from the configuration reached on the previous task. We then evaluate the inherited, adapted and fixed (unedited) configurations on the same \nseqeval\ held-out streams of 3,000 steps, shared by all protocols at that switch, which gives $L_k^{\rm before}$, $L_k^{\rm after}$ and $L_k^{\rm fixed}$. We obtain $\Pi_k$ from equation~(\ref{eq:plasticity}), and Fig.~3b reports $1-L_k^{\rm after}$ and $1-L_k^{\rm fixed}$ averaged over the tasks of each sequence.

\section*{Supplementary Note 7: NISQRC with the Hamiltonian ansatz}
Following ref.~\cite{Hu2024}, we consider $M=2$ memory and $R=4$ readout qubits that evolve for a time $\Delta t$ under
\begin{equation}
\begin{aligned}
U(u)&=e^{-i\Delta t(H_0+u H_1)},\\
H_0&=\sum_{i<j}J_{ij}Z_iZ_j+\sum_i\eta^x_iX_i,\\
H_1&=\sum_i\eta^z_iZ_i,
\end{aligned}
\label{eq:nisqrc_h}
\end{equation}
with all 15 couplers, $J_{ij}\sim U[0,1]$, $\eta^x_i=2+2\epsilon_i$, $\eta^z_i=0.5+0.5\epsilon'_i$ ($\epsilon,\epsilon'\sim\mathcal N(0,1)$) and $\Delta t=1$. With the memory state $\rho^{\rm M}_{n-1}$ and the readout qubits reset to $|0\rangle$, step $n$ produces $\rho_n=U(u_n)(\rho^{\rm M}_{n-1}\otimes|0\rangle\langle0|^{\otimes R})U(u_n)^\dagger$. The features are the 16 readout probabilities $x_j(n)=\mathrm{Tr}[(I_{\rm M}\otimes|j\rangle\langle j|)\rho_n]$, and measurement and reset leave $\rho^{\rm M}_n=\mathrm{Tr}_{\rm R}\,\rho_n$. At finite $S$, the features are multinomial frequencies of $S$ shots. In the channel-equalization task~\cite{Hu2024}, symbols $m(n)\in\{-3,-1,1,3\}$ pass through the channel $x(n)=\sum_{k=0}^{7}h_km(n-k)$ with $h=(1,0.18,-0.1,0.091,-0.05,0.04,0.03,0.01)$, and the received signal is $u(n)=x(n)+0.06x(n)^2-0.01x(n)^3+\epsilon(n)$ with $\epsilon(n)\sim\mathcal N(0,10^{-{\rm SNR}/10})$. The readout is a logistic-regression classifier, as in ref.~\cite{Hu2024}, trained on 100 messages of 100 symbols and tested on 100 further messages. QNL starts from each random instance, and its learning loss is the normalized mean-square error (mean-square error divided by the variance of the symbols in the evaluation messages) of a linear least-squares estimate of $m(n)$ from $S=10^5$-shot features, fitted on 25 and evaluated on 15 messages of each stream ($\nstr=4$ streams of 40 messages). For the tuned baseline, we draw $BK=80$ configurations from the distribution above, evaluate each on four streams with the same loss and shot budget, and keep the configuration with the lowest mean loss~\cite{Bergstra2012}. We run eight independent realizations. With these settings, our implementation recovers the ordering of the reference methods in Fig.~3a of ref.~\cite{Hu2024} (naive rounding, logistic regression, split reservoir, connected reservoir at $S=10^5$ and $S\to\infty$, direct channel inverse). The mean error of the random connected instances is higher than the published curve, and the tuned baseline is close to it (main text).

\section*{Supplementary Note 8: NISQRC with the circuit ansatz and hardware execution}
For each input $u$, the seven-qubit chain R--M--R--M--R--M--R, in which readout (R) and memory (M) qubits alternate, applies
\begin{equation}
\begin{aligned}
U(u)&=\Big[W\prod_i e^{-i(\theta^z_i+\theta^I_iu)Z_i}\,e^{-i\theta^x_iX_i/2}\Big]^{n_T},\\
W&=\prod_{\langle ii'\rangle}e^{-iJ_{ii'}(\Delta t/n_T)Z_iZ_{i'}},
\end{aligned}
\label{eq:nisqrc_c}
\end{equation}
with $n_T=3$, $\Delta t=1$ and $J_{ii'}=1$ on the six chain couplers. We implement equation~(\ref{eq:nisqrc_c}) with the Qiskit gates $R_X(\theta^x_i)$, $R_Z(2\theta^z_i+2\theta^I_iu)$ and $R_{ZZ}(2J_{ii'}\Delta t/n_T)$, applying the even couplers before the odd ones; the four readout qubits are then measured and reset. The angles of the reference instance, $\theta^{x,z,I}_i\sim U[a,2a]$ with $a=(\Delta t/n_T)\theta^{x,z,I}_{\min}$ and $\theta_{\min}=(1.0,0.5,0.1)$, are those generated by the random seed of the authors' public code~\cite{Hu2024}. We generate the 100 test messages of $N=20$ symbols at 20\,dB as in that code, with a circular prefix for the channel memory, and use the received signal $u(n)$ without rescaling. For learning, we simulate the chain exactly in the eight-dimensional Hilbert space of the three memory qubits, using the reset structure of NISQRC. The model includes amplitude damping and dephasing of the memory qubits during every step (from $T_1$, $T_2$ and a step duration computed from the CZ, SX, measurement and reset durations) and, for the readout qubits, relaxation before measurement and symmetric assignment errors. The learning streams use messages generated with seeds disjoint from those of the test messages, and the same variance-normalized surrogate loss as for the Hamiltonian ansatz.
 
On hardware, we rank all seven-qubit paths of the \texttt{ibm\_fez} coupling map whose CZ errors are at most $0.02$ and whose readout qubits have assignment errors of at most $0.05$ by the sum of these errors, excluding qubits flagged as faulty, and select six chains separated by at least one idle qubit. In the first run, each circuit carries three messages, each processed by one NISQRC and one NISQRC+QNL reservoir on its own chain, with NISQRC on chains 1, 3 and 5; 100 messages require 34 circuits of 20,000 shots, preceded by a single 2,000-shot pilot circuit. In the chain-balanced run, circuit $c$ places method $(s+c)\bmod 3$ on chain slot $s=0,\dots,5$ and message $2c+\lfloor s/3\rfloor$. In this way, 50 circuits of 12,000 shots carry each of the 100 messages once per method, and every method uses every chain for 16 or 17 messages. The tuned configuration of this run is the best of 160 configurations (the reference instance and 159 draws from the distribution of the authors), each scored on four fresh streams of 40 messages with the variance-normalized surrogate loss, $S=2^{15}$ shots and the same noise model as QNL; the QNL configuration is the frozen configuration of the first run. We transpile the circuits with optimization level~1, a fixed layout and no routing, and execute them with the Qiskit Runtime sampler~\cite{JavadiAbhari2024} without error mitigation. Mid-circuit measurement and reset are applied as separate operations~\cite{Corcoles2021}, and the frequencies at budget $S$ are computed from the first $S$ shots. Following ref.~\cite{Hu2024}, we average the error rate over 20 random partitions of the 100 messages into 50 training and 50 test messages, with multinomial logistic regression (Newton conjugate-gradient solver, L2 regularization, inverse strength $10^5$) on the 16 readout frequencies of every time step and no washout. The logistic-regression baseline uses $u(n)$ alone with the same protocol. In addition, for the chain-balanced run, we compute the error rate of every message averaged over the partitions in which it is a test message; we group these per-message rates by chain (Fig.~5b) and compare them between methods with paired two-sided Wilcoxon signed-rank tests over the 100 messages.

\paragraph{Noise-aware learning.} We learn the QNL configuration in simulation, using an exact density-matrix model of the chain that includes energy relaxation and dephasing of the memory qubits during each step, relaxation and assignment errors of the readout qubits, and multinomial sampling with $S=2^{15}$ shots. The model uses median parameters typical of current IBM processors ($T_1=253\,\mu$s, $T_2=160\,\mu$s, readout error $6.4\times10^{-3}$), which differ from the calibration of the qubits eventually used, and omits gate errors and crosstalk. QNL edits the coupler set and strengths and the three rotation angles of every qubit, and learns from messages disjoint from the test messages. Over 40 blocks, the surrogate loss decreases from $0.050$ to $0.030$ (means of the first and last five blocks), and the model predicts error rates of $0.045$ (NISQRC) and $0.0070$ (NISQRC+QNL) at $S=2^{15}$; for the tuned configuration of the chain-balanced run, it predicts $0.027$. The frozen configuration uses five of the six chain couplers, and each message requires 600 instead of 720 CZ gates at the same depth (Supplementary Table~\ref{stab:hardware}a).

\paragraph{First hardware run.} In the first run, we place three NISQRC and three NISQRC+QNL reservoirs on fixed, disjoint chains, using 34 circuits of 20,000 shots each (680,000 shots) without error mitigation. QNL reduces the error rate at every shot budget, from $0.248\pm0.010$ to $0.080\pm0.009$ at $S=2^{10}$ and from $0.097\pm0.009$ to $0.022\pm0.004$ at $S=20{,}000$, and falls below the memoryless logistic-regression baseline ($0.053\pm0.006$) from $S=2^{12}$. However, the two methods occupy different chains and no offline-tuned configuration is included, so this comparison cannot separate the effect of QNL from chain-to-chain differences or from the gains that offline tuning could provide. To address this, we perform the chain-balanced run described above.

\paragraph{Comparison with the noise model.} The hardware results differ quantitatively from the noise model. All three configurations perform worse on the device (at $S=12{,}000$) than predicted by the model (at $S=2^{15}$), in line with the gate errors and crosstalk absent from the model, and the QNL configuration shows the largest deviation from the model. The ratio of the tuned-baseline to the QNL error rate is $3.8$ in the model ($0.027/0.0070$) and $1.4$ on the device at $S=12{,}000$, and the ratio of the reference to the QNL error rate is $6.4$ in the model ($0.045/0.0070$) and $3.1$ on the device. At equal shot budgets, the absolute error rates of the same configurations differ between the two runs by up to approximately 35\% in either direction, which is likely due to the different chains and calibrations and indicates the scale of run-to-run variation. The NISQRC error rate on \texttt{ibm\_fez} is higher than that reported in ref.~\cite{Hu2024} on a different processor, where the $ZZ$ interactions were implemented with native echoed cross-resonance gates. In our implementation, these interactions are decomposed into CZ gates, and the reference circuit is not re-optimized for this processor. The tuned baseline shows that part of the gain over the reference can be obtained by offline search in the same noise model; QNL still attains a lower error rate on the device. With a single chain-balanced run on one processor, the variation of this margin across devices and calibration cycles remains to be established.

\begin{table}[t]
\centering\sffamily\footnotesize
\caption{\textbf{Circuits and qubits used on \texttt{ibm\_fez}.} \textbf{a}, Circuit resources after transpilation to the native gate set (CZ, RZ, SX, X), optimization level 1, fixed layout and no routing: one reservoir processing one message of $N=20$ symbols, and the full circuit submitted in the chain-balanced run, which carries six reservoirs (two per method) on disjoint chains. Single-qubit gates include the virtual RZ gates given in parentheses. $^{a}$Estimated as the two-qubit depth times the median CZ duration plus $N$ times the measurement and reset durations. \textbf{b}, Physical qubits of the six seven-qubit chains of the chain-balanced run (roles R--M--R--M--R--M--R), with the CZ error over the six chain couplers, the readout assignment error of the four measured qubits and the median $T_1$ and $T_2$ over the seven qubits, from the backend calibration retrieved with the run. The method-to-chain assignment rotated across circuits, and each chain processed 16--17 messages with each method.}
\label{stab:hardware}
\renewcommand{\arraystretch}{1.15}\setlength{\tabcolsep}{2pt}
\begin{tabular}{@{}lccccccccc@{}}
\multicolumn{10}{@{}l}{\textbf{a}\enspace Circuit resources}\\
\toprule
 & Qubits & Couplers & CZ & Single-qubit (RZ) & Mid-circuit & Resets & Depth & Two-qubit & Duration$^{a}$ \\
 & & & gates & gates & measurements & & & depth & ($\mu$s) \\
\midrule
NISQRC, one reservoir & 7 & 6 & 720 & 4,820 (2,980) & 80 & 80 & 1,420 & 240 & 69.5 \\
NISQRC tuned, one reservoir & 7 & 6 & 720 & 4,820 (2,980) & 80 & 80 & 1,420 & 240 & 69.5 \\
\rowcolor{qnlteal!8}
NISQRC+QNL, one reservoir & 7 & 5 & 600 & 4,183 (2,562) & 80 & 80 & 1,420 & 240 & 69.5 \\
Submitted circuit, six reservoirs & 42 & 34 & 4,080 & 27,646 (17,044) & 480 & 480 & 1,420 & 240 & -- \\
\bottomrule
\end{tabular}

\vspace{8pt}
\setlength{\tabcolsep}{4pt}
\begin{tabular}{@{}ccccc@{}}
\multicolumn{5}{@{}l}{\textbf{b}\enspace Calibration of the qubits used in the chain-balanced run}\\
\toprule
Chain & Physical qubits & CZ error, median (max) & Readout error, & $T_1$ / $T_2$ \\
 & & ($10^{-3}$) & range ($10^{-3}$) & median ($\mu$s) \\
\midrule
1 & 22--23--24--25--37--45--46 & 1.97 (2.38) & 2.9--6.8 & 148 / 132 \\
2 & 137--147--146--145--144--143--142 & 2.04 (2.35) & 2.8--6.0 & 144 / 111 \\
3 & 110--109--118--129--130--131--132 & 2.20 (3.85) & 3.2--5.4 & 166 / 135 \\
4 & 50--51--58--71--70--69--68 & 2.72 (4.89) & 3.5--5.2 & 134 / 88 \\
5 & 54--55--59--75--74--73--79 & 2.35 (2.50) & 3.5--8.7 & 126 / 112 \\
6 & 84--85--86--87--88--89--90 & 2.32 (4.17) & 4.4--7.9 & 143 / 87 \\
\bottomrule
\end{tabular}
\end{table}

\section*{Supplementary Note 9: One-step-ahead forecasting}
\begin{figure}[t]
\centering
\includegraphics[width=\linewidth]{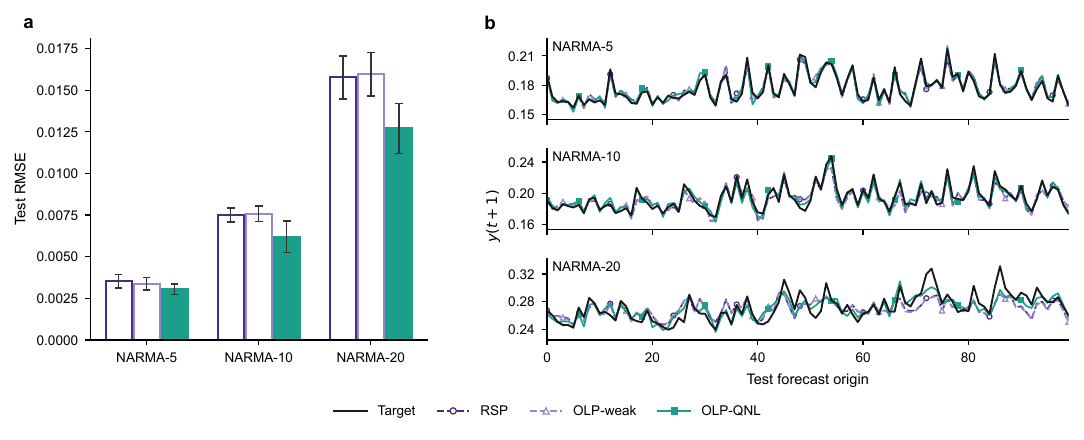}
\caption{\textbf{One-step-ahead nonlinear forecasting.} Five-qubit erase-and-write reservoirs with RSP, OLP-weak and OLP-weak after QNL (OLP-QNL) predict the next NARMA-$n$ output, $y(t+1)$, from the reservoir features at step $t$. \textbf{a}, Test root-mean-square error (RMSE) for $n=5$, 10 and 20; bars, mean $\pm$ s.d. over four paired realizations, 1,000 test forecast origins each. \textbf{b}, Target and predictions for the first 100 test origins of one realization, fixed in advance.}
\label{sfig:forecast}
\end{figure}

For Supplementary Fig.~\ref{sfig:forecast}, we drive the five-qubit erase-and-write reservoir by $s_k\sim U[0,0.2]$ and train a readout of the 31 features at step $t$ to predict $y(t+1)$ of the NARMA-$n$ series (equation~(\ref{eq:narma})) for $n=5$, 10 and 20. The readout standardizes the features with training-segment statistics and fits an unpenalized intercept and ridge weights (penalty $10^{-6}$ per training sample); we use the same readout for all methods. We compare RSP, OLP-weak and OLP-weak with QNL (OLP-QNL), all starting from the same base reservoir. QNL uses the settings listed in Supplementary Table~\ref{stab:settings} ($K=4$, $\nstr=8$, $B=10$, $\delta'=0.05/(KB)$, $\alpha=0.002$), and its learning loss is the validation mean-square error of the forecast divided by the variance of the training targets, on learning streams of 1,100 steps (200 washout, 500 training and the remaining origins for validation, separated by one origin). Next, we freeze the adapted configuration and evaluate it on an independent stream with 1,000 washout, 1,000 training and 1,000 test origins, with one origin between the training and test origins. We compute the test RMSE over the 1,000 test origins for four realizations that share their input streams across methods.

These experiments test whether plasticity improves prediction when the target is the next output of a nonlinear autoregressive series (Supplementary Fig.~\ref{sfig:forecast}). From the features at step $t$, the reservoirs predict $y(t+1)$ of NARMA-$n$ for $n=5$, 10 and 20, using a ridge readout on standardized features, and QNL is trained on this forecasting loss. In Supplementary Fig.~\ref{sfig:forecast}a, OLP-QNL attains the lowest mean test RMSE at every order, and the reduction in RMSE increases with the order. For NARMA-5, OLP-QNL reaches $(3.06\pm0.32)\times10^{-3}$, compared with $(3.52\pm0.41)\times10^{-3}$ for RSP and $(3.38\pm0.37)\times10^{-3}$ for OLP-weak, a reduction of 10--13\%. For NARMA-10, the corresponding values are $(6.19\pm0.95)\times10^{-3}$, $(7.51\pm0.43)\times10^{-3}$ and $(7.58\pm0.46)\times10^{-3}$ (17--18\%), and for NARMA-20, $(12.7\pm1.5)\times10^{-3}$, $(15.8\pm1.3)\times10^{-3}$ and $(16.0\pm1.3)\times10^{-3}$ (19--20\%). Because the two fixed reservoirs perform similarly, the improvement appears to arise from reorganization. With four realizations, the $\pm$s.d. ranges of OLP-QNL and the fixed reservoirs overlap at NARMA-5 and NARMA-10 and separate at NARMA-20, and we do not assign statistical significance to these differences. Because $y(t+1)$ is determined by the inputs up to step $t$, the remaining error reflects the limited memory and nonlinearity of the features. Further, all three reservoirs underestimate the largest excursions of NARMA-20, which depend on inputs up to 20 steps back (Supplementary Fig.~\ref{sfig:forecast}b). These results agree with the memory results of the main text: the benefit of plasticity increases with the memory length required by the task.

\section*{Supplementary Note 10: Statistical procedures}
For Fig.~2 and Supplementary Figs.~\ref{sfig:ad} and~\ref{sfig:shots}, the independent units are reservoir realizations ($n=\nreal$), each with its own base reservoir, learning streams and evaluation streams. For Fig.~3, the independent units are whole task sequences ($n=\nseq$), and for Supplementary Fig.~\ref{sfig:forecast}, four paired realizations. We do not treat points along a curve as independent replicates. We report the results as means with sample standard deviations and 95\% Student-$t$ confidence intervals, together with the paired mean gain between each reservoir and its QNL counterpart; in Supplementary Fig.~\ref{sfig:forecast}, the bars show mean $\pm$ s.d. In Fig.~4, the boxes and bars summarize $n=8$ realizations. For the hardware runs, the error bars give the s.d. over 20 random train/test partitions of the 100 messages and reflect the variability of the data split. In the chain-balanced run, we average the error rate of every message over the partitions in which it is a test message, compare these per-message rates between methods with paired two-sided Wilcoxon signed-rank tests~\cite{Wilcoxon1945} over the 100 messages, and report the per-chain values as means $\pm$ s.e.m. over the 16--17 messages of each chain.

\end{document}